\documentclass{article}
\usepackage[utf8]{inputenc}
\usepackage{graphicx}
\usepackage{color}
\usepackage{url}
\usepackage[toc]{glossaries}
\usepackage{tabularx}
\usepackage{subcaption}
\usepackage{soul}
\makeglossaries

\newacronym{ap}{AP}{access point}
\newacronym{pn}{PN}{pseudo-noise}
\newacronym{prbs}{PRBS}{pseudo-random binary sequences}

\newacronym{thz}{THz}{terahertz}
\newacronym{mmwave}{mmWave}{millimeter-wave}
\newacronym{gscm}{GSCM}{geometry-based stochastic radio channel model}
\newacronym{rt}{RT}{ray tracing}
\newacronym{ral}{RL}{ray launching}
\newacronym{wg}{WG}{working group}
\newacronym{subwg}{subWG}{sub-working group}
\newacronym{ris}{RIS}{reconfigurable intelligent surfaces}
\newacronym{isac}{ISAC}{integrated sensing and communications}
\newacronym{jsac}{JSAC}{joint sensing and communications}
\newacronym{los}{LoS}{line-of-sight}
\newacronym{3d}{3D}{three dimensional}
\newacronym{3gpp}{3GPP}{3rd generation partnership project}
\newacronym{sage}{SAGE}{space-alternating generalized expectation–maximization}
\newacronym{mimo}{MIMO}{multiple-input multiple-output}
\newacronym{ue}{UE}{user equipment}
\newacronym{bs}{BS}{base station}
\newacronym{iot}{IoT}{internet of things}
\newacronym{drt}{DRT}{dynamic ray tracing}
\newacronym{ml}{ML}{machine learning}
\newacronym{ai}{AI}{artificial intelligence}
\newacronym{tdl}{TDL}{tapped delay line}
\newacronym{cdl}{CDL}{clustered delay line}
\newacronym{mpc}{MPC}{multipath component}
\newacronym{sl}{SL}{supervised learning}
\newacronym{ul}{UL}{unsupervised learning}
\newacronym{rl}{RL}{reinforcement learning}
\newacronym{dt}{DT}{decision tree}
\newacronym{gbdt}{GBDT}{gradient boosting decision tree}
\newacronym{dl}{DL}{deep learning}
\newacronym{mlp}{MLP}{multi-layered perceptron}
\newacronym{cnn}{CNN}{convolutional neural networks}
\newacronym{ofdm}{OFDM}{orthogonal frequency division multiplexing}
\newacronym{csi}{CSI}{channel state information}
\newacronym{elaa}{ELAA}{extremely large antenna array}
\newacronym{ds}{DS}{delay spread}
\newacronym{as}{AS}{angular spread}
\newacronym{pdp}{PDP}{power delay profile}
\newacronym{ple}{PLE}{path loss exponent}
\newacronym{rms}{RMS}{root mean square}
\newacronym{2d}{2D}{two dimensional}
\newacronym{fft}{FFT}{fast Fourier transform}
\newacronym{ioe}{IoE}{Internet of Everything}
\newacronym{crlb}{CRLB}{Cramér-Rao Lower Bound}
\newacronym{em}{EM}{Expectation-Maximization}
\newacronym{eadf}{EADF}{Effective Aperture Distribution Function}
\newacronym{music}{MUSIC}{MUltiple SIgnal Classification}
\newacronym{esprit}{ESPRIT}{unitary Estimation of Signal Parameter via Rotational Invariance Techniques}
\newacronym{dmc}{DMC}{dense multipath component}
\newacronym{dss}{DSS}{directional scanning scheme}
\newacronym{dked}{DKED}{double knife-edge diffraction}

\title{\textbf{White Paper on Radio Channel Modeling and Prediction to Support Future Environment-aware Wireless Communication Systems}}
\author{Editors: Mate Boban and Vittorio Degli-Esposti 
}
\date{September 2023}

\begin{document}
\begin{figure}[!t]
	\centering
	\includegraphics[width=0.36\columnwidth]{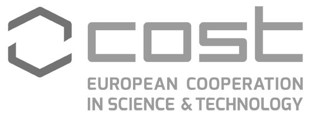}
 \hspace{2.5cm}
 \includegraphics[width=0.4\columnwidth]{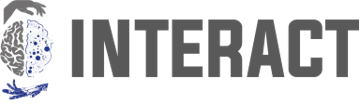}
\end{figure}


\maketitle

\vspace{2cm}

\begin{center}
\Large
COST CA20120 INTERACT \\
Working Group 1 (Radio Channels)    
\end{center}
\normalsize

\vspace{2cm}

\begin{figure}[htbp]
	\centering
 \includegraphics[width=0.4\columnwidth]{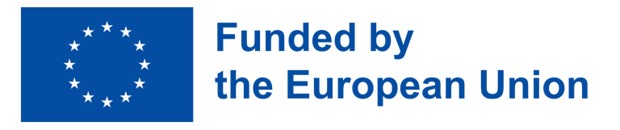}
\end{figure}

\newpage

\glsunsetall
\tableofcontents

\glsresetall
\newpage
\printglossary[title=Acronyms, toctitle=Acronyms]



\newpage
\section{Introduction}
\hspace*{8.5mm}\textbf{by Mate Boban}\\

\begin{figure}[!t]
	\centering
	\includegraphics[width=\columnwidth]{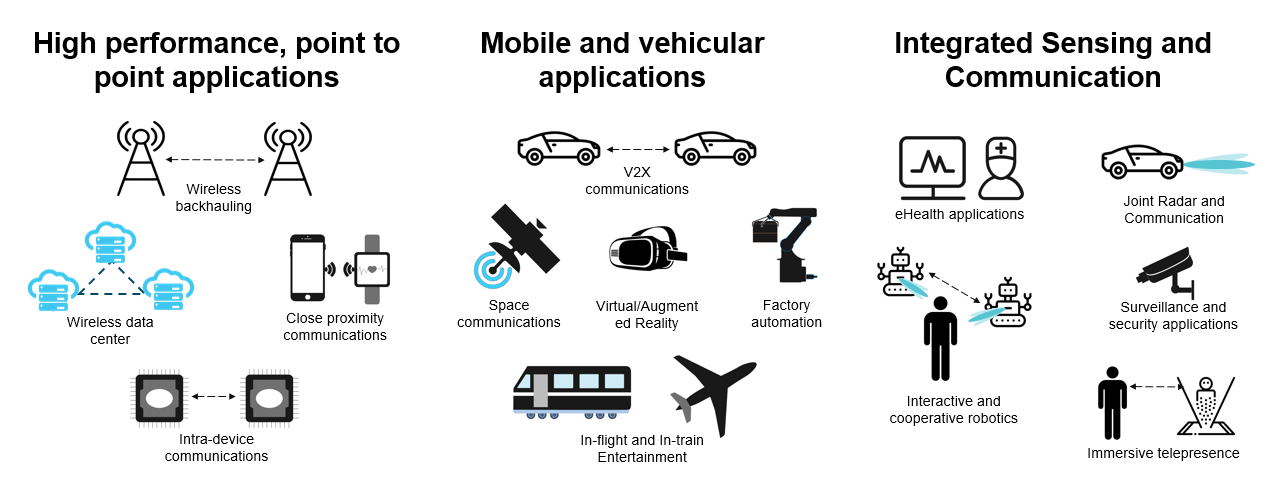}
	\caption{Selected use cases that next generation wireless communications systems aim to enable (based on~\cite{one6G}).}
	\label{useCases}
\end{figure}

COST INTERACT \gls{wg}1 aims at 
increasing the theoretical and experimental understanding of radio propagation and channels in environments of interest and at deriving models for design, simulation, planning and operation of future wireless systems. Wide frequency ranges from sub-GHz to \gls{thz}, potentially high mobility, diverse and highly cluttered environments,  dense networks,  massive antenna systems, and the use of intelligent surfaces, are some of the challenges for radio channel measurements and modeling for next generation systems. As indicated in~\cite{salous22}, with increased number of use cases (e.g., those identified by one6G~\cite{one6G} and shown in Fig.~\ref{useCases}) to be supported and a larger number of frequency bands, a paradigm shift in channel measurements and modeling will be required. To address the particular challenges that come with such a paradigm shift, \gls{wg}1 started the work on relevant topics, ranging from channel sounder design, metrology and measurement methodologies, measurements, modeling, and systematic dataset collection and analysis. 

In addition to the core activities of \gls{wg}1, based on the strong interest of the participants, two \glspl{subwg} have been initiated as part of \gls{wg}1: i) \gls{subwg}1.1 on \gls{mmwave} and \gls{thz} sounding (\gls{subwg} \gls{thz}) and ii) \gls{subwg}1.2 on propagation aspects related to \gls{ris} (\gls{subwg} \gls{ris}). 

This white paper has two main goals: i) it summarizes the state-of-the-art in radio channel measurement and modeling and the key challenges that the scientific community will have to face over the next years to support the development of 6G networks, as identified by \gls{wg}1 and its \glspl{subwg}; and ii) it charts the main directions for the work of \gls{wg}1 and \glspl{subwg} for the remainder of COST INTERACT duration (i.e., until October 2025).

In this white paper, particular attention has been devoted to the concept of ``environment awareness'', which is defined as 
the ability of communications systems to sense in the broad context of object detection, positioning and ranging, and even object imaging. The repercussions of environment awareness are therefore discussed throughout the paper, ranging from the  definition of use cases and their requirements (Section~\ref{sec:Environments}), to required frequency bands (Section~\ref{sec:Bands}), to the considerations on the  design of channel measurements (Section~\ref{sec:Measurements} and definition of channel  models (Section~\ref{sec:Methodologies}), finally to definition of new technologies such as \gls{isac} (Section~\ref{sec:NewTechnologies}).

\subsection{Mandate of \gls{wg}1}

\begin{figure}[!t]
	\centering
	\includegraphics[width=\columnwidth]{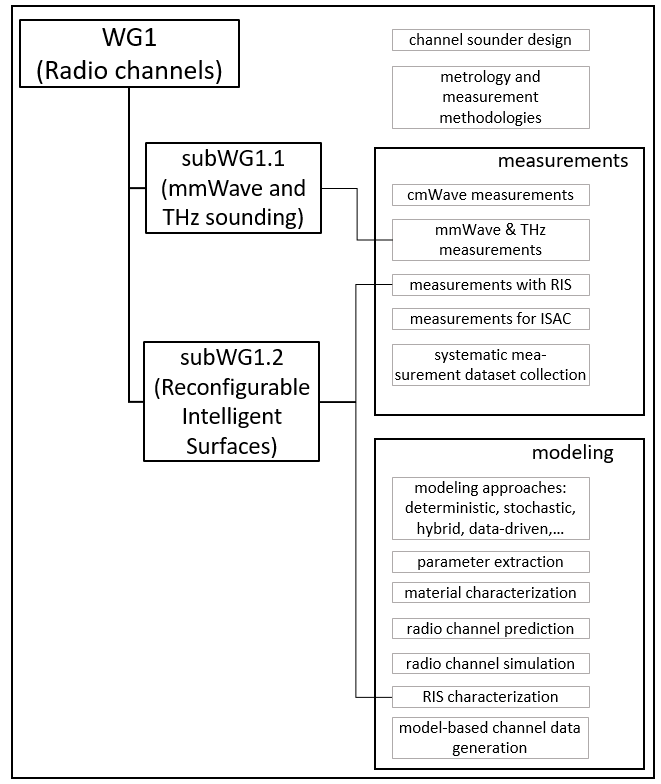}
	\caption{Topics addressed by \gls{wg}1 and its \gls{subwg}s. 
 }
	\label{topics}
\end{figure}

Extensive efforts are being  devoted to obtaining a comprehensive understanding of radio wave propagation in several frequency bands for the development of future wireless networks. The task of \gls{wg}1 is to further this understanding by providing an open and collaborative forum for the exchange of ideas, definition of key challenges, and identification of directions for research on radio channels. To that end, the efforts in \gls{wg}1 relate to propagation modeling for radio systems, including the ones exploiting \gls{mmwave} and higher frequency bands (sub-\gls{thz} and \gls{thz}), where large contiguous bandwidths are still available, and massive \gls{mimo} and beamforming techniques, which will enable spectral-efficient connectivity in densely populated areas. Understanding radio wave propagation has also been crucial to new applications, including highly dynamic scenarios, \gls{iot} and smart grids. Efforts on propagation modeling for these systems and applications encompassed vehicular and \gls{mmwave} cellular access~\cite{salous2016millimeter}, \gls{iot}, Smart Grids~\cite{sandoval2017improving}, and energy efficient cellular radio planning. These propagation modeling studies have been carried out using various measurement setups~\cite{cardona2016cooperative} or combining measurements and theory for link- and system-level simulations~\cite{wu2017general}, addressing the time, angle and polarisation characteristics of multipath channels, as well as the characterisation of material properties, outdoor-to-indoor penetration loss, and link blockage~\cite{salous2016millimeter}. Propagation models become mature once supported by a vast amount of measured and simulated evidence of radio channels, and by our understanding about them. Ultimately, such mature propagation models may be able to perform real-time prediction of radio environments and hence provide accurate-enough \gls{csi} to aid radio communication systems and applications. As an example, a few studies have addressed the real-time use of deterministic propagation models to help estimate \gls{csi} \cite{fuschini2019study}, along with a location-aware \gls{csi} fingerprinting~\cite{di2014location}. Their real-time use in localisation, beamforming, and resource allocation algorithms is still in its infancy.

\gls{wg}1 is also committed to collecting data and sharing them to create large reference sets for model development and training of \gls{ml} approaches, with \gls{wg}1 members already contributing several datasets for this purpose (datasets available at \url{https://interactca20120.org/wgs/datasets-2/}).

Based on the discussion above, Figure~\ref{topics} summarizes the main topics that \gls{wg}1 is addressing, along with the indication of specific topics that are handled by the \glspl{subwg}. \gls{wg}1 will contribute to each of these topics in order to reach its ultimate goal --  definition of a comprehensive channel modeling framework that addresses new scenarios and frequency bands proposed for future wireless communications systems.


\subsubsection{Mandate of \gls{subwg}1.1}
The goal of the \gls{subwg}1.1 (\gls{mmwave} and \gls{thz} sounding) is to concentrate the expertise on radio channel measurements and analysis related to mm-wave frequencies and up, which represent the frontier in experimental channel characterization. Novel experimental set-ups, verification of channel sounders, radio channel measurements in different environments/for different applications are some of the key aspects to be investigated. The chain to be covered ranges from the validation of the measurement equipment to the analysis of the measurement results. The objectives of this \gls{subwg} are to extend the knowledge on propagation from empirical analysis and to develop common practices, in order to create a rich pool of harmonized data from diverse sets of measurements.
\subsubsection{Mandate of \gls{subwg}1.2}

The goal of the \gls{subwg}1.2 (\gls{ris}) is the analysis and modeling of propagation in future smart radio environments empowered by controllable and smart surfaces. These surfaces enable the manipulation of propagation characteristics, including wavefront shape and polarization, and the minimization of signal losses. This is particularly of interest at \gls{mmwave} and higher frequencies to extend otherwise limited communication ranges due to high path losses and the very poor signal penetration into the shadow regions of objects and obstructions. Taking the idea further, large numbers of smart surfaces would enable smart environments where it is possible to optimize the channels and to maximize the network throughput and efficiency. 

In order to understand the benefits and limitations of radio channel modulation with \gls{ris}, we have to properly understand the modeling and performance of different types of \gls{ris} (e.g., reflect arrays, metasurfaces, holographic surfaces, etc.). Modeling those properly is a key to continue to the modeling and analysis of individual links, entire systems, and smart environments. Therefore, the goal of this \gls{subwg} is to extend the knowledge on three fundamental aspects of \gls{ris}: 1) proper and realistic \gls{ris} models for different \gls{ris} configurations and technologies, 2) propagation models for \gls{ris} empowered links and systems, and 3) performance of smart links, systems, and environments.

  

\subsection{White paper organization}

For each subsequent section of the white paper covering technical topics, we attempted to cover the following: i) a brief description of the state of art, including key references; ii) summary of COST INTERACT \gls{wg}1 contributions to the topic\footnote{While COST INTERACT contributions are not made public, vast majority of contributions are published in conferences, journals, or other venues, either before or after submission to COST INTERACT. Throughout the paper, we refer to those publicly available versions of contributions.}; and iii) identification of future work needed on the topic. By implementing this approach, we hope the white paper serves as a reference for the researchers looking to get a primer on channel measurements and modeling for future communications systems.

The rest of the white paper is structured as follows. Section~\ref{sec:Environments} describes the relevant environments and channel modeling scenarios and Section~\ref{sec:Bands} discusses the frequency bands of interest. Section~\ref{sec:Framework} provides an introduction into the wireless channel propagation fundamentals. Relevant channel sounder designs and metrology are covered in~\ref{sec:Sounder}. Based on the identified environments and frequency bands,  Section~\ref{Sec:sounding} discusses the channel measurements. Channel modeling methodologies are covered in Section~\ref{sec:Methodologies}, and
Section~\ref{sec:Estimation} discusses channel parameter estimation. 
Section~\ref{sec:NewTechnologies} identifies new technologies and techniques related to channel modeling that need to be implemented in order to properly evaluate future communications systems. Finally, Section~\ref{sec:Conclusions} provides an outlook on the future work 
and concludes the white paper.

\newpage
\setcounter{footnote}{0} 
\section{Scenarios}\label{sec:Environments}

\hspace*{8.5mm}\textbf{by Vittorio Degli Esposti }\\


Early generations of wireless networks were conceived for a limited number of propagation environments and use-cases. In terms of physical characteristics, environments were classified into rural, urban and indoor~\cite{3gpp.25.943}, which corresponded to an increasing attenuation and traffic density, and therefore to a decreasing cell radius. Further classification into suburban and dense-urban, or large-indoor, office and residential indoor is also widely used. The most important use case was that of voice services and internet access connectivity using mobile \gls{ue} such as smart phones, tablets and laptop computers.
Over the years, driven by technology advances and market demand, wireless networks have evolved into a multi-technology integrated galaxy of systems, with a large variety of connected devices, scenarios and propagation environments, as depicted in Fig.~\ref{useCases}. 
Besides the traditional scenarios described above, novel scenarios will include the use of new frequencies in the THz and optical bands and of densified, cell-free networks in high-traffic areas, the realization of the “Network of Everything” with massive connectivity of objects and machines to realize “Smart Environments” and “Smart Factories”, also with the use of ubiquitous Artificial Intelligence and of Reconfigurable Intelligent Surfaces, the realization of \gls{3d} networks including drones and UAVs as network components, and finally the implementation of automated and connected cooperative driving scenarios using dedicated or cellular-based networks and of Joint Sensing and Communication techniques .


Although the concept of ``environment'' or ``scenario'' is a vague one that encompasses physical – and therefore propagation – characteristics, frequency band, technology solutions and applications, here we provide a brief overview of the wide variety of scenarios that next-generation systems will likely have to address, with reference to the classification in Fig.~\ref{useCases} and with particular emphasis on COST INTERACT WG1 research.
\paragraph{Indoor and in-X high-performance-link scenarios.}
It is well known that 6G-and-beyond systems will have to raise the bar of achievable performance in term of transmission speed, throughput density, low latency and reliability. This will require the use of higher frequency bands in the mm-wave, \gls{thz} and optical ranges, and of network densification \cite{one6G, 9040264, 8901159}. 
The possibility of very high bitrates - of the order of Tbps - and very low latencies of \gls{thz} links will enable new application scenarios for indoor and very short-range communications - also known as in-X communications - such as high-definition holographic infotainment and "teleportation", ultra-broadband mobile access for offices and public spaces, high-performance wireless communication links for industrial applications, data centers, in-vehicle  inter-device and intra-device connections \cite{8901159}.
Visible Light Communications (VLC) are envisioned especially for indoor environments, where illumination LEDs, already strategically deployed across indoor premises, can be conveniently reused for communication \cite{9208801}.

\paragraph{Vehicular scenarios.}
Future transportation scenarios will be characterised by high mobility and will involve cars, trains and unmanned aerial vehicles flying at low altitudes. All of them will require massive use of radio applications including Vehicle to Vehicle (V2V), Vehicle to Infrastructure (V2I) and Vehicle to Everything (V2X) connections as well as radar and \gls{isac} schemes to ensure cooperation, control and safety \cite{8410403}. The dynamic nature of radio links and networks in the transportation scenarios is the key feature to be addressed in radio channel modeling research.
Within the foreseen automated and connected cooperative driving application scenarios, key assets are a reliable wireless V2X connectivity and accurate localization. Most recent vehicles and last-generation wireless systems users will have such capabilities, while others will not (heterogeneous traffic). Environment-aware and cooperative solutions will have to take advantage of connectivity, localization and mapping information available to the former kind of road users or to the edge cloud, to enforce safety for the latter and the whole traffic system.
Cooperative, Passive Coherent Radar solutions are being proposed where signal emitted by the fixed infrastructure and vehicles equipped with V2X can be reused as multi-static radar sources that the system can opportunistically use to determine the location of vehicles and pedestrians along streets and in proximity of road intersections \cite{8847233}.

\paragraph{\gls{isac} and environment awareness.} 
The \gls{csi} or at least information about multipath spatial characteristics should be known at both radio-link ends to fully exploit the potential of massive \gls{mimo} schemes as well as of highly directive mm-wave and \gls{thz} links in mobile environments. The problem will therefore have to be addressed, for example through \gls{isac} techniques, \gls{ai} techniques \cite{yang2023AI}, or real time use of digital-twin schemes with embedded propagation models \cite{9711524, fuschini2019study}. \gls{isac}, especially at \gls{thz} frequencies, will enable high-definition environment "vision" applications, including environment mapping, medical imaging, surveillance applications, safety enhancement applications in vehicular applications, etc. All these methods, combined with the ubiquitous use of \gls{ai} can be thought as enablers of the so called “Environment Awareness” and "intelligence" of future systems. 

\paragraph{Smart and Reconfigurable Environments.} 
\gls{ai}, universal wireless connectivity (\gls{ioe}) and \gls{ris} technology will enable Smart and Reconfigurable Environments. For the first time in the history of wireless systems, \gls{ris} and the use of Unmanned Aerial Vehicles or drones (\gls{3d} networks) will allow the customization of the propagation environment with the purpose of optimizing performance and enhancing the application potential \cite{DiRenzo2020SmartRadio, 9781659}. \gls{ris} allow the manipulation of the reflected or transmitted wavefront, enabling interesting applications such as anomalous (non-specular) reflection, focalization, signal processing. Properly placed and configured \gls{ris} can be used to enhance mm-wave and \gls{thz} coverage, or to optimize channel capacity or system performance with a so-called “Scatter \gls{mimo}” approach \cite{DiRenzo2020SmartRadio, 9781659}. The combination of \gls{ris} and UAV is also interesting in highly cluttered mm-wave and \gls{thz} scenarios to compensate for blocked \gls{los} communication links and create controllable and smart radio environments \cite{9781659, 9367288}.
\paragraph{Industrial Environment.} 
A particular candidate to become "smart" is the Industrial Environment where advanced wireless networks will allow a variety of disruptive applications. Cable replacement with ultra reliable and high-performance wireless links is very attractive for the great flexibility and increased reliability of connections with sensors/actuators in moving parts and with robots. The use of mm-wave and sub-THz frequencies will allow ultra low-latency (below 0.1 ms) connections to avoid oscillating behaviours in control loops while enabling high-definition environment sensing through \gls{isac} to control the production process and enforce safety \cite{9390169, tan2021integrated}.

\section{Frequency bands of interest}\label{sec:Bands}
\hspace*{8.5mm}\textbf{by Mate Boban}\\

Spectrum is the main consideration for each generation of wireless communication technology as more spectrum is needed to support higher data rates \cite{tong20216G}. Since mobile communication technologies evolve to new generations, the use of spectrum continues to expand to higher frequency bands. The spectrum expected to support environment-aware communications can be divided into the following frequency bands.
\begin{itemize}
    \item \textbf{Sub-6 GHz band}: Virtually all of the spectrum up until 4G (LTE) has been allocated in the sub-6 GHz band. This band continues to play a crucial role in 5G and is expected to be vital in 6G as well. This frequency band is the most cost-effective option as a frequency range to guarantee wide coverage in mobile communication systems.
    \item \textbf{Mid-band}: The frequency bands between approximately 6-24~GHz are also identified as competitive candidates for supporting the continued growth of traffic and environment-aware applications, especially given the better sensing performance of larger array size per unit area.  To support the continuous growth of traffic, at least 1 to 1.5 GHz of additional spectrum is needed. The 5925-7215 MHz range has been identified as a potential candidate to provide the needed spectrum. Moreover, compared to the sub-6GHz band, the propagation attenuation of these bands increase in an acceptable range while path loss will be combated by adopting advanced radio technologies, e.g., massive \gls{mimo} \cite{8352131}.  
    \item \textbf{\gls{mmwave} band}: The \gls{mmwave} band contains a relatively large amount of available bandwidth, which is essential for ultra-high data rates and high-accuracy sensing applications. However, operation in the \gls{mmwave} band is more challenging due to the unfavorable propagation characteristics compared to lower frequency bands. In the 2015 world radiocommunication conference (WRC), a variety of frequency ranges were proposed for IMT sharing study between 24 and 86 GHz and in WRC 2019; specifically a total of 17.25 GHz was identified \cite{tripathi2021millimeter}. E-bands (71-76 and 81-86 GHz) are prime candidates to support larger contiguous blocks in the future, mainly reserved for non-geostationary fixed-satellite service systems (space-to-earth and earth-to-space) \cite{acts2012world}. The upper and lower 60 GHz, namely the 57-64 GHz and 64-71 GHz frequency ranges, further provide large contiguous chunks of bandwidth to support device-to-device communications through access and backhaul links and aeronautical and land mobile services, respectively \cite{tripathi2021millimeter}. Furthermore, technologies such as integrated access and backhaul (IAB) could make use of the available spectrum available at the \gls{mmwave} band \cite{9040265}.
    \item \textbf{Sub-\gls{thz} and \gls{thz} bands}: Sub-\gls{thz} and \gls{thz} bands open new possibilities for sensing and communication \cite{THzMagazine}. A total of over 100GHz in 92-275 GHz band is \textit{allocated}, whereas 130 GHz in 275-450 GHz band is \textit{identified} for mobile services or land mobile services~\cite{studies2019WRC}. At these frequencies, there are several parts of contiguous spectrum exceeding 10 GHz, which makes it possible to support very high data rates for short- and medium-distance communication. In addition, \gls{thz} bands bring enhanced sensing resolution thanks to the ultra-wide bandwidth and shorter wavelength \cite{9376324}. Further up the frequency, between 450~GHz up to 10~\gls{thz} there is potential for further spectrum. While there exists an unprecedented amount of spectrum in these bands, they also experience new challenges~\cite{THz2021}: extremely high transmission losses, molecular absorption that creates non-monotonic pathloss over different frequencies, variability due to weather conditions, effect of micro-mobility, etc. These effects need to be addressed to ensure efficient use of the large spectrum. Additionally, the use of visible light spectrum~\cite{pathak2015visible} has gained significant momentum in recent years, given its ability to support novel communication and sensing use cases.
\end{itemize}

\begin{figure}
\begin{center}    
  \includegraphics[height=1.5cm,width=11cm]{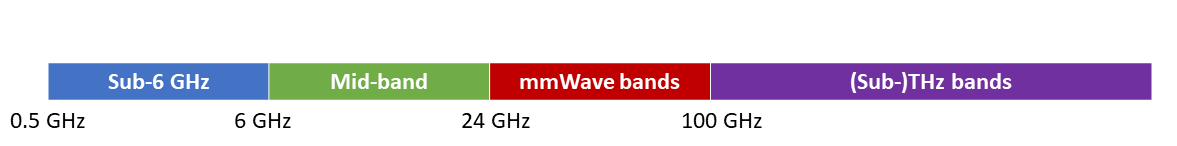}
\end{center}
\caption{Spectrum bands.}
\label{fig:freqz_band}  
\end{figure}

\newpage
\section{Wireless channel propagation fundamentals} \label{sec:Framework} 

\hspace*{8.5mm}\textbf{by Vittorio Degli Esposti, Conor Brennan, and Katsuyuki Haneda}\\

\subsection{Basic propagation mechanisms} 
\label{subsec:propMech}

While an exact description of radio propagation and therefore of a radio channel might be theoretically possible via Maxwell's equations, it would be unwieldy and complex. Instead, it is possible to simplify propagation description - while retaining relevant features and characteristics - in terms of a set of basic mechanisms that will be reviewed in this section.  

\subsubsection{ Waves in homogeneous media} 
The simplest case to consider is free-space, or propagation in vacuum.   In this case
the power density decays inversely with the square of the distance, essentially reflecting the fact that a constant amount of total source output power is being  spread over a larger total surface area as the wave propagates further away from the transmitting antenna.   For communication systems there is an additional loss, proportional to the square of the frequency, due to the frequency dependence of the effective aperture of a receiving antenna.  At the high mm-wave and THz frequencies of interest within COST Action CA20120 this is a significant issue but can be mitigated by the use of narrowly focused beams at transmitter and receiver.  However this relies on the continued existence of clear line of sight path.
This is explored in \cite{TD(22)03045} in which high resolution dual-polarised double directional measurements are taken in the context of  industrial control communications operating at 300GHZ.  Propagation and blockage spatial and temporal characteristics are obtained from the processed data indicating the presence of viable alternative communication paths.

Propagation within any other homogeneous material that can be found in a radio channel (e.g. water, air, building materials, human tissue) is  broadly similar in many respects to the free space case. The precise physical effect of a particular material can be described with reference to its constitutive parameters, namely its electric permittivity, magnetic permeability and conductivity. These parameters capture the macroscopic effects of the material’s atomic and molecular structure on any waves passing through them. 
These effects manifest in several ways including a change in the wave’s phase velocity (slowing, relative to the speed in vacuum) and also a change in the characteristic impedance (the ratio of the amplitude of the electric and magnetic fields) and the wavelength (the physical distance between successive peaks or troughs along the wave).  Importantly,  the presence of conductivity or dielectric hysteresis in so-called lossy media manifests itself as an extra reduction in the power density as the wave propagates (in addition to the spreading discussed previously).
These effects are frequency-dependent which leads to dispersion effects as the individual frequencies which comprise a pulse travel at different speeds through the material, causing the pulse to distort.

\subsubsection{Reflection, transmission, diffraction and scattering }
\begin{figure}[!t]
	\centering
	\includegraphics[width=\columnwidth]{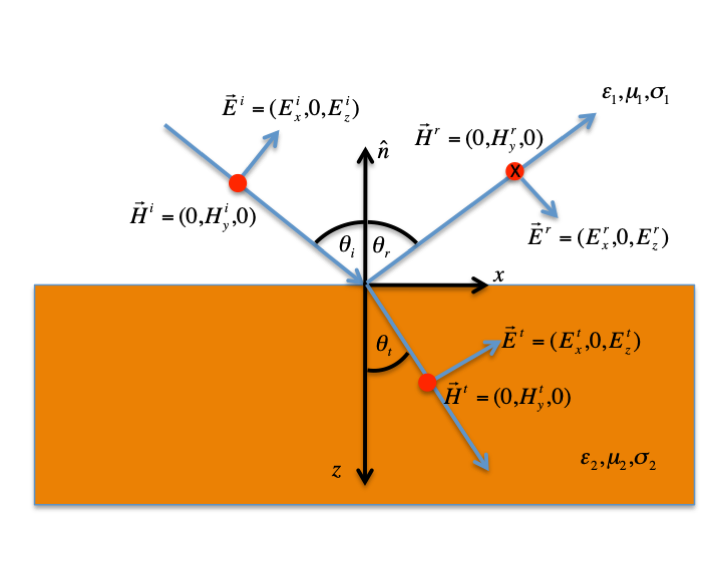} \caption{Plane wave reflection from smooth boundary. Incident wave ${\vec E}^i, {\vec H}^i$ produces a reflected wave  ${\vec E}^r, {\vec H}^r$ and a transmitted wave  ${\vec E}^t, {\vec H}^t$}
	\label{reflection_example}
\end{figure}

The primary complication afflicting radio propagation is the proliferation of waves   occurring at the {\em boundaries} between materials (such as when a wave travelling in air strikes a wall).  Referring to figure (\ref{reflection_example}) {\bf reflection} and {\bf transmission} occur when an incident electromagnetic wave strikes the face of an object which is locally smooth on a scale comparable to the wavelength. In such circumstances the incident wave produces a reflected wave travelling away from the face and a transmitted wave propagating into the object, the direction of propagation of both being governed by Snell’s laws of geometric optics. 
The amount of power reflected from, and transmitted through, materials is  frequency dependant and their specification for  new communication frequencies at millimetre wave and higher is an important task being addressed in COST Action CA20120. Several contributors have conducted studies of reflection and transmission (penetration) loss for typical building materials, concentrating on their variation with angle, frequency and polarisation. \cite{TD(22)01012} described the use of a wideband channel sounder to examine a variety of materials at four discrete frequencies between 28~GHz and 70~GHz and noted the increase of penetration loss with frequency.   Continuous measurements of reflection and transmission losses are made in \cite{TD(22)02007} and \cite{TD(22)02008} for 17 common materials in the frequency range 2GHz to 170GHz.  Oscillations in the reflection loss as the frequency varies are noted, due to the effect of internal reflections within the finite slab of material under test. These diminish as the frequency rises, consistent with the increased penetration loss as noted elsewhere.    

Another research focus is on the development of so called reconfigurable intelligent surfaces \gls{ris} which do not obey the above laws of geometric optics but rather can reflect or transmit signals in preferred directions, thereby improving coverage or reducing interference.    
Within COST Action CA20120 researchers are   examining ways to accurately model them using commercial Finite Element software \cite{TD(22)02066}.  This numerically intensive approach is shown to give good agreement with the simpler Generalised Law of Reflection approach but has the added advantage of being able to model scattering in all directions and thus can be used for interference analysis. More details on COST Action CA2 20120's work on RIS are available in     section~\ref{sec:RIS}

\begin{figure}[!t]
	\centering
	\includegraphics[width=\columnwidth]{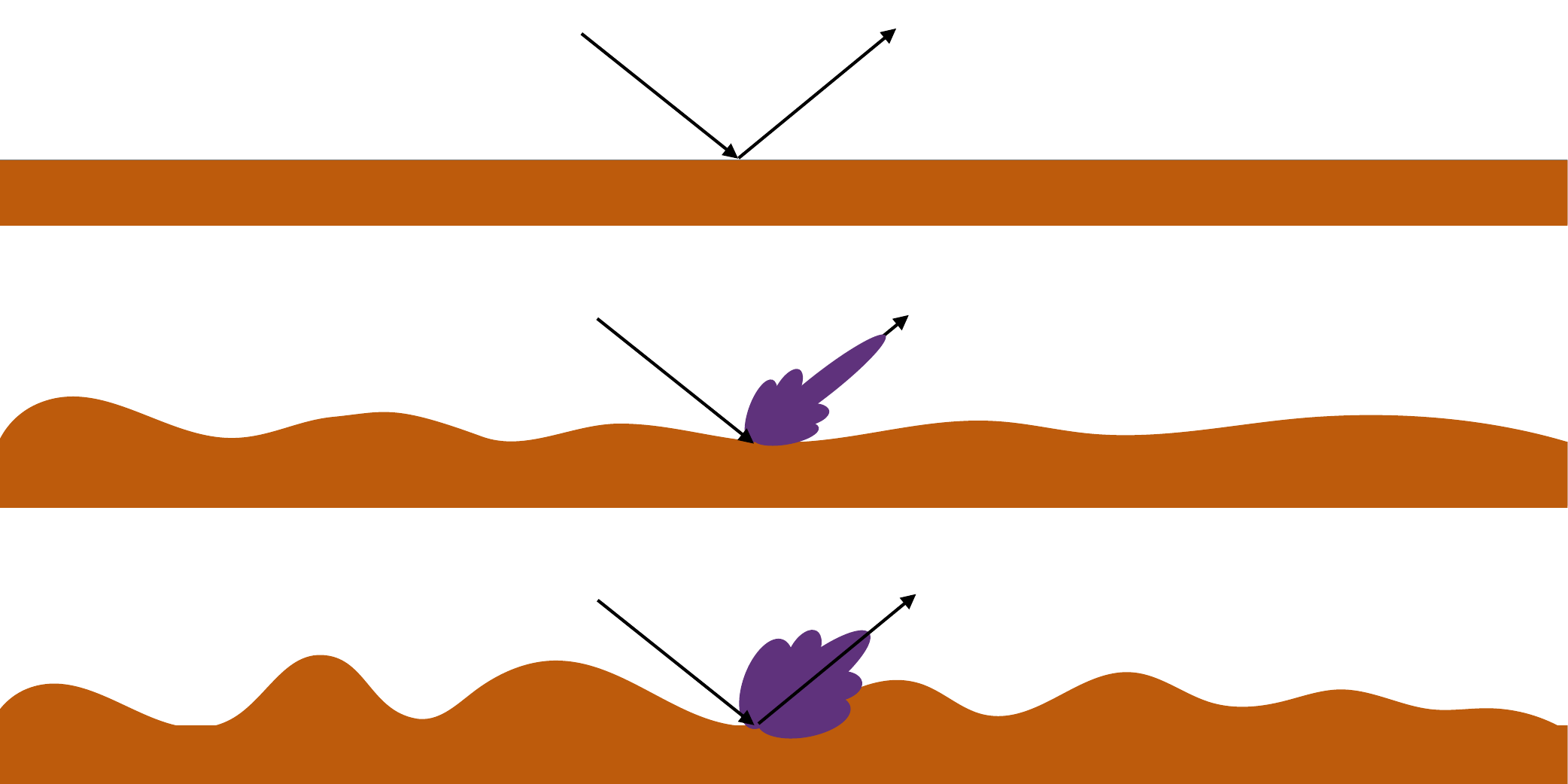} \caption{Laws of geometric optics  gradually give way to diffuse scattering (reduced specular lobe and increased side lobes)   as surface roughness increases}
	\label{diffuse_example}
\end{figure}

Diffraction occurs when a wave strikes the sharp boundary between two such faces (such as at the edge of a building). In such instances the wave is scattered in a continuum of directions, as defined by the so-called Keller cone \cite{Balanis}. As the interface between regions becomes rougher (or equivalently the frequency becomes higher such as is the case with \gls{mmwave} and THz communications) finer details  at wavelength scales become important and these simple well-defined mechanisms of reflection, transmission and diffraction give way to the more general process of {\bf diffuse scattering}, which as the name suggests results in a proliferation of waves being scattered diffusely across a wider angular range (as per figure (\ref{diffuse_example}).  {\bf Surface scattering}, where the inhomogeneities are assumed to lie on the material boundary, has been studied by participants in  COST Action CA20120 with a variety of models proposed. \cite{TD(22)01045} and \cite{TD(22)01091} both use numerically precise models based on the method of moments with the former paper concentrating on efficient ways to solve the associated computationally intensive equations while the latter contribution examined depolarisation effects and the dependency on material, frequency and surface size.    The accuracy of a simple directional scattering (DS) model is assessed in \cite{TD(22)02010} and effective roughness and scattering coefficient are identified as key parameters.   The model is extended using a $t$ location-scale distribution in order to make it better fit electromagnetic simulation results. 

Several authors have identified the  process of {\bf volume scattering}, which occurs when waves interact with fine inhomogeneities {\em within} a material, as being of more importance than surface scattering. The Effective Roughness model is a relatively simple tunable model and has been used to describe such scattering at mm-wave frequencies \cite{TD(23)04053}. In this contribution it was noted that materials with complex internal structure (such as reinforcing mesh) can produce significant backscattering resulting in a diffuse scattering model  involving two directional lobes (in the forward and backward directions). The effective roughness model is heuristic but is based on a solid insight of the physical processes that take place at a material boundary. In \cite{TD(22)03055} the authors enhance its scientific basis by modifying it so that it obeys the reciprocity principle (i.e. is unchanged by an interchange of transmitter and receiver locations).  
As seen, diffuse scattering models are often statistical or heuristic in nature, given the uncertainty surrounding the physical form of  small scale inhomogeneities that produce them.  This uncertainty is caused by an unavoidable limitation in the level of detail of  databases describing buildings etc.    Nonetheless some COST INTERACT contributors have examined what can be achieved with an enhanced level of building geometric description. In  \cite{TD(22)03054} the impact of building facade detail is studied by comparing ray tracing output (see below) to measured data. It is concluded that the inclusion of enhanced information about facade features (windows, ledges etc) can result in a more discriminating identification of multipath components than would be the case with simple diffuse scattering models applied to flat facades.

\subsection{Large-scale and small-scale propagation phenomena}

As discussed in Section~\ref{subsec:propMech}, the quality of a radio channel is ultimately determined by how electromagnetic waves interact with the materials within it. Recognising that such a full electromagnetic description is practically impossible, channel modeling has instead focused on describing the channel in terms of a number of key parameters that have the greatest effect on the performance of the digital communication scheme being enabled by the channel. The range of parameters and the accuracy with which they need to be specified have evolved in tandem with the communication schemes themselves as they grown in sophistication over the decades.   Parameters are divided into large scale parameters and small scale parameters, depending on the range over which they vary significantly compared to the wavelength.

\subsubsection{Large scale propagation phenomena}

\begin{figure}[!t]
	\centering
	\includegraphics[width=\textwidth]{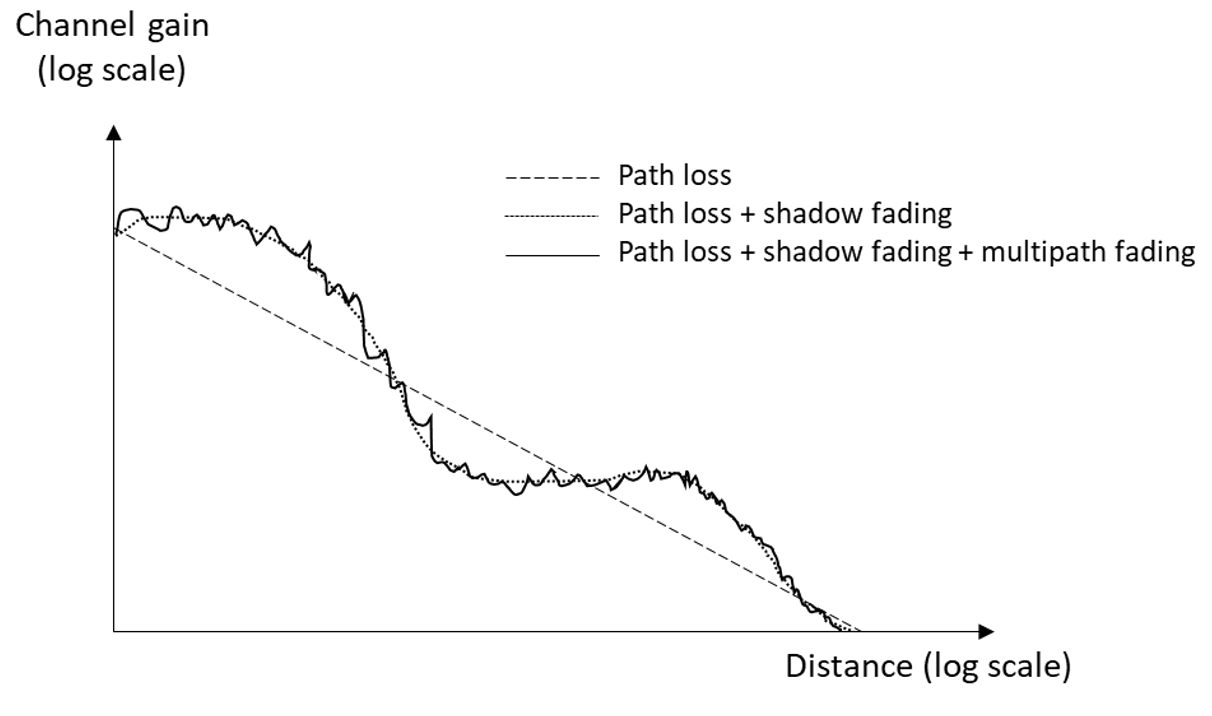}
	\caption{Path loss, shadow fading, and small scale/multipath fading.}
	\label{PL_SF}
\end{figure}

\textbf{Path loss}\\
Path loss describes the large-scale reduction in power density as signals move away from the transmitter. It is the most important parameter in channel modeling since it ultimately determines the possible communications range and its quality, i.e., the signal-to- noise ratio. It also determines the signal-to-interference ratios (SINR) in the case of an interference-limited environment within which frequencies are reused to ensure efficient use of spectrum. From an energy-efficient networks perspective path loss defines the minimum transmit power of base and mobile stations required to maintain a target SINR and quality of service. Consequently, path loss is the most well-studied parameter in channel modeling. Path loss is typically modeled using a power law, wherein the loss is proportional to the distance raised to the power of some specified exponent.   Path loss exponents typically range from 1.5 to 4 for waveguiding to deep non-line-of-sight environments. Frequency dependency of the path loss has also been studied thoroughly, indicating good agreement with the theory, e.g., Friis’ law with constant-gain antennas. Even though it is a subject that has been extensively studied, it remains a subject of research when new spectrum, use cases and scenarios emerge, e.g., rural and aircraft scenarios. \\
Referring to the log-log plot of Fig.~\ref{PL_SF}, path loss is typically depicted as the best linear fit extracted from the measurement data.

\textbf{Shadow fading}\\
Path loss models assume a regular decay of power density with distance (e.g. linear or piecewise linear when viewed on a log scale), In practice the power density will display further random large-scale variations around this trend as some locations suffer from a greater or lesser degree of obstruction by buildings etc. This is referred to as {\em shadow-fading} and is typically modeled using a suitable random distribution. 
In legacy below-6GHz cellular radio systems, shadow fading is typically defined by fluctuation of signal strength over different sectors of \glspl{bs} at the same communication distance between base and mobile stations. A single \gls{bs} site consists of multiple \gls{bs} units, each covering a limited azimuth and/or elevation angular range called a sector. The level of obstruction in each sectoral direction varies, with some possibly experiencing large blockages with a consequent decrease of signal strength. The standard deviation of shadow fading is therefore not necessarily identical to the second order moment of differences between path loss and small-scale-averaged received signal strength normalized to the transmit power. Still, such a definition of shadow fading is the only one that can be used for radio communications systems and scenarios where the use of sectors would not be popular, e.g., in indoor scenarios. An open research question of shadow fading is its correlation over space and frequency~\cite{Burhal19_Globecom}. 
Referring to Fig.~\ref{PL_SF}, shadow fading adds a slowly varying process on top of the path loss.
\\

\textbf{Blockage}\\
Blockage refers to the obstruction of the \gls{los} path between a transmitter and a receiver by an object (e.g., building, wall, human body, etc.).
It can be modeled either as part of the shadowing fading process or explicitly, where the latter is preferable in case of a known/measured additional blockage loss by a specific object~\cite{du2018suburban, virk2019modeling}. In case of explicit modeling, it is added as an additional loss on top of path loss and shadow fading. Below we elaborate in particular on human blockage, as it becomes increasingly relevant for handheld or on-body user devices utilizing \gls{mmwave} or higher frequency spectrum.  


It has been shown through experiments that losses of a \gls{los} connectivity due to human blockage is reproduced well by modeling a human body as a set of absorbing knife edges and considering diffraction on each edge, where the \gls{los} path is totally absorbed by the absorbing screen representing the body. The use of absorbing knife edges is advantageous to model human body in its simplicity because diffraction coefficients do not depend on polarization of an incident wave, in contrast to the case of conducting screens. As the analytical formula of diffraction coefficients assume infinitely long edges, they are better applicable to higher radio frequencies where the human body becomes electrically larger. A research question would be estimation of blockage loss when there are {\it multiple} objects intervening the \gls{los} connectivity~\cite{Ramos17_AWPL, Prado-Alvarez21_CL}. The analytical treatment becomes much more complex than a single body case because diffraction coefficients of an absorbing knife edge assume incident plane wave, while an incident wave to the following edges after the first one may not be a plane wave necessarily. The same problem encounters when estimating extra losses to free-space due to multiple diffraction over hills and buildings in long-range point-to-point links. In addition, as the cross sectional area of the Fresnel zones becomes larger as longer connection distance, estimation of the blockage losses are more challenging because only a part of the Fresnel zones may be intervened by blocking objects.

\subsubsection{Small-scale propagation phenomena}

Large scale effects such as path loss and shadow fading describe variation of power density that occur gradually on a scale of many wavelengths. In practice it is noted that power density varies rapidly on the scale of the wavelength also. This phenomenon, called \emph{fast fading},is caused by wave interference between EM waves arriving at a given location via multiple paths of differing lengths, i.e. by the propagation mechanism usually referred to as \emph{multipath}.  Given the impossibility of precisely specifying these path lengths this is an intrinsically random process and a detailed statistical modeling of such small-scale fading effects is therefore an essential part of modern radio channel models. Fast-fading signal-strength fluctuation are described using a variety of statistical distributions, Raileigh and Rice distributions being the most widely used for non-\gls{los} and \gls{los} channel conditions, respectively.
Multiple parameters are used to model time, frequency and space dispersion effects of multipath propagation. These effects and are often represented by Fourier transform pairs, e.g., space/angle, time/Doppler and delay/frequency, the choice of which domain to use depending on the channel sounding and modeling methods. 

Channel models usually prescribe the second moments of the power spectrum in the respective domains, i.e., angular, delay and Doppler spreads, or their Fourier counterparts, i.e., spatial, frequency and time correlation intervals.  These parameters have been studied thoroughly through a wide range of channel sounding from a few hundreds of Megahertz to sub-\gls{thz} radio frequencies across many important radio communications scenarios. However, it must be noted that the sole spread or correlation parameter values do {\em not} suffice to reproduce channels that resemble realistic conditions because many different shapes of power spectrum or correlation functions yield the same parameter values. Explicit knowledge, i.e., shapes of power profiles and correlation functions is therefore usually accompanied with parameter values, as discussed in the following.\\

\textbf{Multipath dispersion and clustering}\\
\label{sec:mf}
Power spectrum of multipaths over angles and delays typically does not have equal magnitude over the domain, but shows some extents of power concentration on specific angles and delays in each radio link. Such concentration of power in the spectrum is represented by {\it clusters} in multipath channel modeling. Each cluster can be defined by its concentration angles and delays, which are called cluster centers, along with their distributions in the respective domain. The latter is typically modeled by Laplace and exponential distributions in the angular and delay domains. Setting the right cluster centers and types of distributions allows us to reproduce the realistic power spectrum shapes of a radio link realization while respecting their spread parameter values. The use of clusters for multi-dimensional power spectrum modeling of multipath channels is a well-established approach, as evidenced by standard channel models that rely on them. An open research question is the variation of angular and delay properties of clusters across different radio frequencies, especially at \glspl{mmwave} and higher frequencies. It has been discussed among the wireless communication community that multipath channels are sparser, i.e., the number of multipaths and/or clusters is smaller, at higher radio frequencies. There are some indications from comparative channel sounding performed at various radio frequencies~\cite{Virk17_EuCAP, Nguyen18_EuCAP, Dupleich19_EuCAP, Dupleich19_Access} that support the conjecture, where the power spectrum becomes more ``spiky", i.e., dominated by specular reflections while reduced scattering effects, so that clusters become more distinguishable to each other.  The clearer specular reflections and reduced scattering are both explained by wave-interacting objects whose surface roughness becomes comparable to the wavelength of the radio frequency, making the wave-object interaction more angularly selective and the number of multipaths arriving at the receive side smaller becomes less. The same reports tend to show that, despite the sparsity, the spread parameter values do not necessarily change noticeably. Some other reports, on the other hand, show evidence from channel sounding at multiple frequencies that the power spectrum shapes do not change noticeably across frequencies~\cite{Vehmas16_VTCF, Mate19_Access}. The discussion of multipath sparsity and their influence on spread parameter values is therefore not conclusive yet, requiring further evidence from multiple-band channel sounding in different scenarios. They have implications on cluster models across the radio frequencies in standard channel models.
Referring to Fig.~\ref{PL_SF}, multipath fading results in fast variations superposed on top of the path loss and shadow fading.

\newpage
\section{Channel sounder design \& metrology}\label{sec:Sounder}
\begin{itemize}
    \item[] \textbf{by Diego Dupleich and Wei Fan}
\end{itemize}

\subsection{Channel Sounding Design}

Channel sounding consists of ``sounding'' the environment with a known signal and analysing the received echoes to characterize the propagation of electromagnetic waves. Since both the transmitted and the received signal are known, the time-variant channel impulse response can be extracted. The complexity of the sounding set-up depends on the propagation parameters under investigation and target channel models and systems. Nowadays, with mobile wideband MIMO systems at high frequency in sight, the ultimate goal of multidimensional channel sounding is to provide the necessary data to jointly estimate the amplitude (polarization) and different geometrical properties of the multi-path components in the channel: delay, direction of departure (DoD), direction of arrival (DoA), and Doppler. In practice, the simultaneous acquisition of these \gls{mpc} properties is challenging and there is frequently a trade-off between resolution in the delay domain (bandwidth), angular domain (directivity), and Doppler (sampling rate).

\subsubsection{Wideband Channel Sounding}

Channel sounders are usually classified as frequency-domain or time-domain channel sounders.

\begin{figure}[tb]
	\centering
	\subfloat{%
			\includegraphics[scale = 0.6]{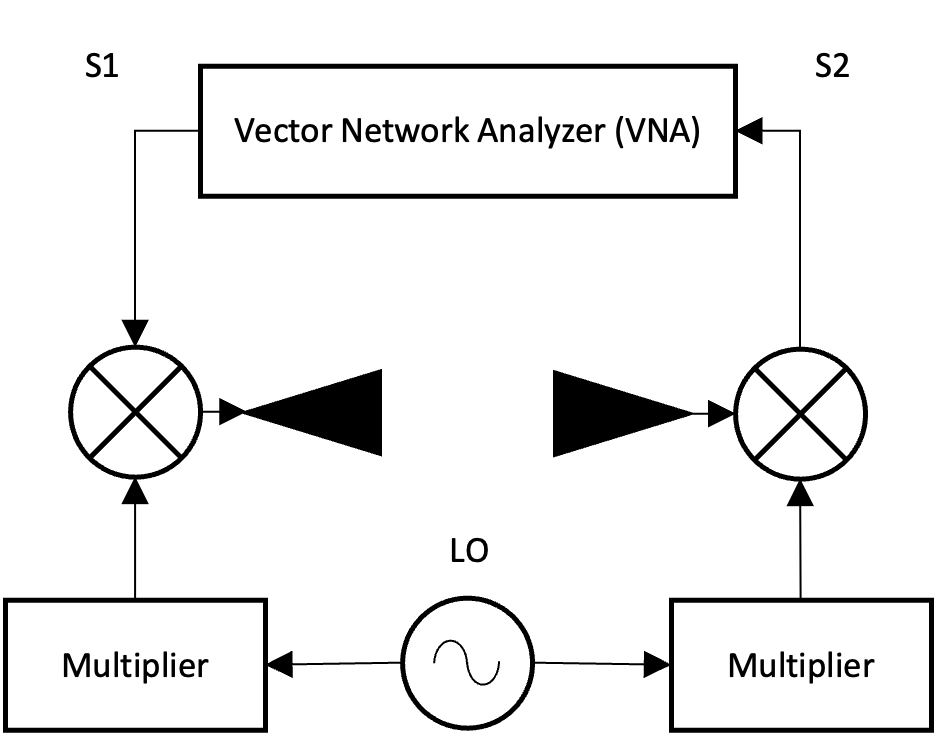}}  \\
	
 (a)
	
	\subfloat{%
			\includegraphics[scale = 0.6]{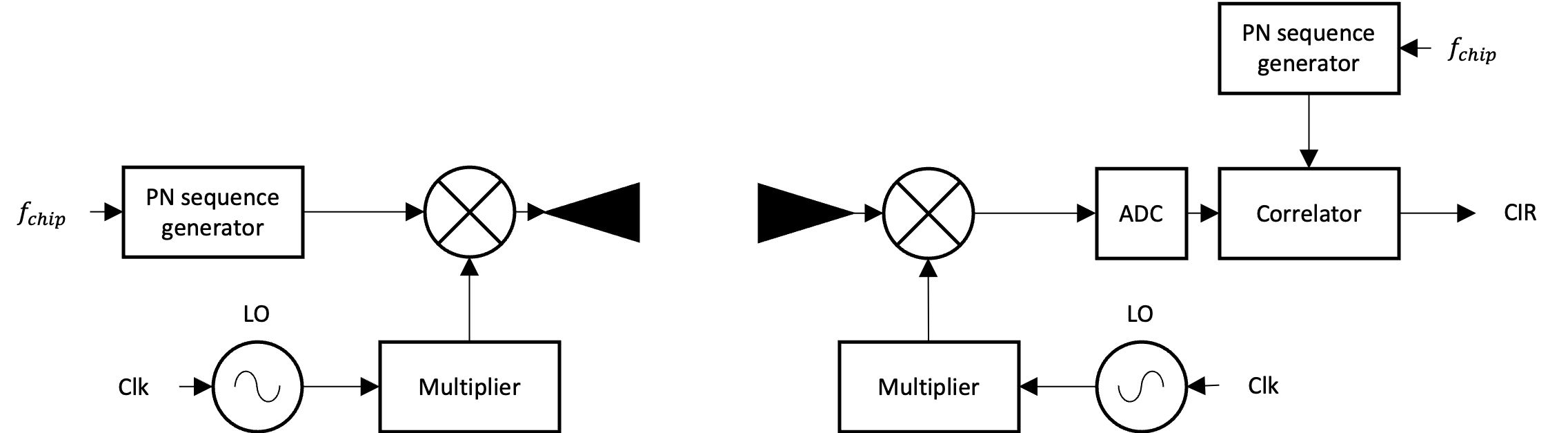}} \\

 (b)
	\caption{Wideband channel sounding architecture (a) VNA-based and (b) correlation based. 
 }
	\label{fig:channel_sounding_architecture}	
\end{figure}

Vector network analyser (VNA) is a widely used type of
frequency-domain channel sounder, recording the frequency response between two ports of the device, as shown in Fig.~\ref{fig:channel_sounding_architecture}a. 
They are popular due to the ease of calibration, excellent dynamic range, scalable and flexible carrier frequency and bandwidth settings. With the help of external frequency extenders, it supports channel measurements in the \gls{mmwave}/(sub-)THz bands as well. However, there are some shortcomings, for example, slow measurement time (determined by the number of swept frequency points and IF settings), short measurement distance (due to high losses in the RF cable used to remote antennas), and lack of VNA ports to support multi-antenna measurements. However, different solutions have been investigated in INTERACT to work around some of these limitations. Radio-over-fiber (RoF) has been implemented to solve the problem of short measurement range for \gls{mmwave}/(sub-)THz bands \cite{AAU_longrange_1,AAU_longrange_2}. Moreover, a phase-compensation scheme that counteracts random phase variations has been tested in \cite{AAU_sub_THz_phase_coherent_1,AAU_sub_THz_phase_coherent_2}, achieving accurate and phase coherent measurements for VNA-based \glspl{dss} and virtual-array schemes at \gls{mmwave} and (sub-)THz.   

On the other hand, time-domain channel sounders utilize specially designed wideband signals. Popular signals are multi-carriers and \gls{pn} / \glspl{prbs} (displayed in Fig.~\ref{fig:channel_sounding_architecture}b), which have special auto-correlation properties in time. In contrast to multi-carrier, PRBS have the advantage of a low peak-to-average power ratio (PAPR), allowing the optimization of transmit power, a scarce resource at high frequencies. On the other hand, unlike with multi-carrier signals, the spectral power density of PRBS is not uniform over the measured bandwidth, which also creates difficulties on the frequency response calibration of the system. In comparison to VNAs, time-domain channel sounders allow real-time measurements. However, this requires expensive wideband digitizers that can be saved by implementing sub-sampling or the sliding correlation architecture. This also has the advantage of an increased SNR at the expense of larger measurement times.

\subsubsection{Antennas and Antenna Arrays for \gls{mmwave} and (sub-)THz Channel Sounding}

For \gls{mmwave}/(sub-)THz applications, it is of high importance to have knowledge of the spatial profile of the channel. Several strategies have been reported in the literature to capture this information, including the \gls{dss} and the utilization of virtual and physical antenna arrays. 
In the \gls{dss} scheme, a highly directional antenna can be centred on a turntable and rotated to record the spatial profile of the channel. The \gls{dss} scheme has been the most popular solution for \gls{mmwave} and (sub-)THz bands because of its simplicity, low cost and high-gain offered by the employed antenna. However, it is slow due to the mechanical steering nature and its spatial resolution is essentially constrained by the directivity of the antenna \cite{AAU_longrange_1,AAU_longrange_2}. The virtual array solution is another popular scheme that employs a single antenna which is sequentially moved in pre-defined spatial locations (i.e. virtual elements of the array) \cite{AAU_sub_THz_phase_coherent_1,AAU_sub_THz_phase_coherent_2}. However, the \gls{dss} and virtual array schemes can only be used for static propagation scenarios due to long measurement time. On the other hand, time-varying scenarios can be measured with physical arrays. The multiplexing antenna array scheme consists of several directional antennas pointing to different spatial directions connected to a radio frequency (RF) switch, enabling the direct recording of the spatial profile \cite{AAU_massive_MIMO_channel_sounder}. To decrease the hardware complexity and cost while maintaining the performance of the full digital structure,  a switched array based channel sounder for mmWave bands was reported in \cite{mmWavesounder}. By activating different antenna pairs, channels of all antenna pairs are measured at different time instants. Switched-array sounding can accomplish one MIMO channel snapshot within a very short measurement time. However, non-sequential antenna switching is needed to overcome the aliasing in the parameter estimation of Doppler frequencies and angles of \glspl{mpc} \cite{hybridSwitching}. With phased arrays (i.e., with an analog beam-forming structure), the process of beam-steering is much faster and it has been widely used in the \gls{mmwave} bands \cite{NIST_phased_array_sounder}. Though highly promising, its application to large-scale configuration at \gls{mmwave}/(sub-)THz bands has not been reported for channel sounding. More flexibility and capabilities compared to phased arrays can be achieved with a digital beam-forming structure, in which each antenna element has an individual RF chain \cite{AAU_massive_MIMO_channel_sounder}. However, the number of RF channels is essentially limited due to cost and complexity concerns. 

\subsection{Metrology of Channel Sounding}

Metrological assessment of multidimensional channel sounding has gained relevance in the latest years as it can be seen from different activities around the world: the METERACOM project\footnote{https://www.meteracom.de}, the NextG Channel Model Alliance\footnote{https://www.nist.gov/ctl/nextg-channel-model-alliance.} (sponsored by NIST),  
and standardization activities in the IEEE P2982 group (\gls{mmwave} Channel Sounder Verification), which is working towards recommendations of methods for verifying \gls{mmwave} channel sounders performance. These verification methods are usually based upon comparison of processed channel measurements to either theory or to particularly designed artefacts 
that generate \glspl{mpc} with known characteristics in the different domains \cite{8658008}. These artefacts can also be measured over-the-air \cite{9769269}, which offers the versatility of including the measurement antennas and allowing mobility during the test, enabling the verification of angle of arrival/departure and Doppler \cite{8904605}.

\subsection{Future Work}

Within COST INTERACT \gls{wg}1 and \gls{subwg}1.1, we will be tackling the reduction of measurement time in virtual-arrays and \gls{dss} schemes. It is important to build statistically meaningful 6G channel models, which requires double directional channel measurements at multiple locations  for many deployment scenarios. However, a key bottleneck for channel sounding at \gls{mmwave}/(sub-)\gls{thz} frequency bands using the DSS and virtual arrays is the long measurement time associated with the mechanical movement of the antenna. Virtual array based channel sounder for the W-band (75~GHz - 110~GHz) was designed and employed for channel measurements in \cite{AAU_sub_THz_phase_coherent_1,AAU_sub_THz_phase_coherent_2}. We will continue working on developing virtual array channel sounder for the 220~GHz to 330~GHz frequency band. A few works on channel sounder to support cell-free massive MIMO scenarios were reported within COST INTERACT \cite{cell_free_massive_MIMO_channel_sounder}, which we will continue to cover. The implementation of physical-arrays at (sub-)THz for dynamic measurements and joint estimation of Doppler/angle/delay of \glspl{mpc} will be addressed from a metrological and practical point of view. Within the scope of \gls{isac}, there is plenty of space for the optimization of baseband excitation signals to enhanced sounding performance, reduce uncertainties during measurements, and optimally exploit the hardware resources. Non-linearities are also challenging with the utilization of wideband multi-carrier signals and need to be addressed from a metrological and practical point of view. 


\newpage
\section{Channel measurements}\label{Sec:sounding}
\begin{itemize}
    \item[] \textbf{by Mate Boban, Diego Dupleich, Wei Fan, Marco Skocaj, and Wenfei Yang}
\end{itemize}

\subsection{sub-6 GHz Frequency Band}
The first four generations of cellular systems as well as IEEE 802.11 (WiFi) standards up until a decade ago were all enabled by sub-6 GHz bands. Therefore, over the last more than 30 years, a large number of measurement campaigns have been carried out to characterize wireless propagation in sub-6~GHz bands~\cite{molisch2005ultrawideband}. 

However, these bands still garner interest, primarily due to emerging application and deployment scenarios and use of novel antenna techniques. To that end, recent studies presented at COST INTERACT meetings focused on the characterization of high mobility scenarios, such as vehicular and indoor factory environments, and considered new deployment approaches, such as cell-free and massive \gls{mimo} architectures.  

In \cite{10001039}, authors conducted channel measurements at 3.2 and 5.81 GHz in vehicular propagation environments, including V2V, V2I, and V2P scenarios. Together with channel measurements, the authors collected LiDAR data captured by sensors installed in the connected vehicles and built a dataset with coherent perception and propagation traces. 
In \cite{simon:hal-03952309}, authors performed measurements in a cell-free vehicle-to-infrastructure communication scenario using a real-time channel sounder operating at 5.89 GHz and with a bandwidth of 80 MHz. At the transmitting side, they considered different setups with different number of access points and antenna elements. At the receiving side, multiple omnidirectional antennas were installed in a van and spaced more than 10 lambdas, enabling the measurement of dynamic and decorrelated channels. For each setup, the authors evaluated SNR and \gls{rms} \gls{ds}, and demonstrated that cell-free architecture can guarantee a better and spatially more uniform link quality.

In \cite{https://doi.org/10.1049/mia2.12244}, authors performed wideband and ultra-wide band measurements and adopted a Bayesian approach to derive and empirical fading model for device-free localization purposes. They considered different setups, including outdoor measurements at 5.2~GHz 
and indoor measurements at 4~GHz. In \cite{9900633}, authors carried out a measurement campaign to characterize indoor-to-outdoor high-mobility propagation scenarios at 5.9~GHz. The transmitter antenna is installed inside a building and placed on a rotating unit that rotates with a constant velocity. The receiver is placed on the roof of another building at a distance of 140~m. 
The results include normalized local scattering functions for different velocities of the transmitted antenna.

In \cite{willhammar2022fading}, authors considered a factory environment and conducted measurements at 3.7~GHz considering two different deployment options: a co-located massive \gls{mimo} antenna array and a unique randomly distributed array. Measurement results are used to quantify the channel hardening effect observed when increasing the number of antennas, and show that the usage of massive \gls{mimo} arrays in rich scattering environment can reduce large scale power variations and help in achieving more reliable wireless channels.

\subsection{Mid-band (6~GHz - 24~GHz)}
As the spectrum below 6~GHz is limited, current services are already using a large proportion of the available spectrum. Therefore, resorting to higher frequency ranges becomes a necessity. 
Frequency bands between 6~GHz - 24~GHz are considered as a promising candidate for supporting future wireless communications systems due to the lower path loss compared to higher (\gls{mmwave} and above) bands and a potential to support massive \gls{mimo} systems.

While up to now there have been limited contributions in terms of measurements in mid-bands by COST INTERACT (e.g.,\cite{midBandELAA}), channel measurements for frequency bands between 6~GHz and 24~GHz have been widely conducted in the literature, where indoor scenarios \cite{CorOffLabConf11n14, Lobby10, RomHallMuse11, Lec15, Hall15, Theater11, ClaMet10} have driven more research interest over outdoor scenarios \cite{resid11, BuildCorn10}.

The study in \cite{midBandELAA} presented measurements for \gls{elaa} with a 32×32 Rx virtual planar array in 10~GHz band in two indoor environments (meeting room and classroom). Results indicate strong non-stationary effects in the spatial domain as well as significant near-field effects, both of which become non-negligible in \gls{elaa} channel modeling. As the antenna arrays become larger 
and since massive \gls{mimo} systems are expected to operate in mid-bands, 
further investigations of non-stationarity are needed 
with \gls{elaa}. 
To that end, in \cite{Hall15}, a 20×20 virtual uniform rectangular array (URA) was used at Rx side in channel measurements at 13–17 GHz in a lecture hall environment. 
Channel parameters, including channel gain, K-factor, and DS were observed over the array which showed considerable variations without deterministic trends. 
In \cite{Theater11}, a 64×4 virtual URA was used at Tx side in channel measurements at 11 GHz in a theater environment and channel parameters such as shadow fading, \gls{ds}, and coherence bandwidth were derived over the array. Similar to~\cite{midBandELAA}, the spatial non-stationarity was again pronounced. 

Primary channel parameters including path loss, shadowing, K-factor, \gls{ds}, \gls{as}, cross-polarization ratio (XPR), and correlation properties have been obtained in several environments. 
In \cite{CorOffLabConf11n14}, channel measurements were conducted at 11~GHz and 14~GHz in indoor environments, where the antenna elements were configured in a 7×7 square grid. 
The \gls{ple} and \gls{ds} were derived from the averaged \gls{pdp}  to remove the effect of small-scale fading. 
\cite{Lobby10} presented a complete parametrization for a three-dimensional (3-D) \gls{gscm} at 10.1 GHz based on a measurement campaign in a lobby environment. 
To obtain the 3-D spatial information, a virtual conformal array consisting of four 9×9 planar arrays was used in the measurements.  
The estimation of signal parameters via rotation invariance techniques (e.g., ESPRIT) were employed to estimate the \gls{mpc}.
\glspl{mpc} were first clustered by the ``Power K-means’’ algorithm and then cluster parameters were also obtained.  
The study in \cite{resid11} carried out similar parametrization procedure based on channel measurement at 11~GHz in a micro-cell environment. 
A virtual uniform circular array (UCA) with 12 elements were used in measurements,  hence the angular spread was only available for the azimuth plane. 
In \cite{Lec15}, measurement campaigns were conducted by the \gls{dss} at 13–17~GHz in an indoor lecture hall and a laboratory environments with high gain horn antennas.  
The \gls{sage} algorithm was applied to de-embed the effect of antenna response and then the \gls{ds} and \gls{as} were obtained. 

The characteristics of specific propagation mechanisms, such as diffuse scattering and diffraction, have been also studied for the mid-band. \cite{RomHallMuse11} characterized diffuse scattering based on \gls{mimo} channel measurements at 11 GHz in indoor environments. Propagation parameters of diffuse scattering were jointly estimated by the RiMAX-based estimator, where incoherent plane waves due to diffuse scattering is modeled stochastically as dense multipath component (DMC).  The measurement results showed that significant DMC exist, which affected the eigenvalue structure of the \gls{mimo} channel. 

In summary, existing channel measurement campaigns in mid-band have been conducted in various indoor but few outdoor environments. The measurement results covered statistical channel parametrization, non-stationarity analysis for \gls{mimo} channels, and propagation mechanisms study. 

Future measurement campaigns in mid-band are expected to cover more propagation scenarios, especially outdoors, which are necessary for defining proper channel models for outdoor scenarios in these bands. Further measurements and analysis are required to explore characteristics of massive \gls{mimo} channels, including how the propagation environment affects massive \gls{mimo} system performance in mid-band.

\subsection{\gls{mmwave} and (Sub-)\gls{thz} band}
The characterization of propagation from measurements at mmWave and the lower THz bands has gained a lot of attention in the recent years. The free blocks of spectrum available in these frequencies enable the implementation of high data rate wireless links with enhanced capacity and with an unprecedented level of resolution that makes them suitable for sensing applications. 

However, there are propagation aspects related to the wavelength at mmWave and (sub-)THz differing to the well known and studied sub-6~GHz bands. As the frequency goes up, transmission loss will become high, diffraction becomes much weaker, and penetration becomes very difficult, making the propagation channel much sparse and specular. In addition, in the mmWave and (sub-)THz range, particles in the atmosphere requires further considerations for some frequency bands.

To compensate the increased isotropical path-loss, high-gain radio interfaces need to be employed. Therefore, the spatial characteristics of the channel become more relevant for channel modelling and system design, since the information on where to point the antenna beam is of special importance. Consequently, even empirical path-loss models have also been adapted from the typical isotropical characteristics of the antennas to consider the directivity \cite{7109864}.



Therefore, there is paradigm shifting from purely stochastic towards hybrid stochastic/deterministic models for these frequency bands with the inclusion of deterministic components. This requires a more precise characterization of the electromagnetic properties of different constructive materials and a deeper analysis on the scattering properties. 
The absorption coefficient and refractive index of typical building materials (glass, plaster, and wood) has shown a good agreement between the measurements and the results of the Fresnel equations, \cite{1572680}. On the other hand, diffuse scattering is modelled by extending Fresnel equations for specular reflections with a Rayleigh factor obtained from the knowledge of statistics of surface roughness by Kirchhoff theory of scattering \cite{4380579}.

COST INTERACT contributions on the topic of material penetration and reflection losses in a wide range of frequencies covering from sub-6 GHz to the lower THz bands (up to 170~GHz) have shown a relative low dependence on frequency of the reflection coefficients for the majority of the materials under test, but an increased penetration loss above 100~GHz \cite{9769465, 9815763}. In addition, measurement-based analysis from different typical construction materials at 27~GHz showed that the scattering from internal structures can be relevant, especially in the case of Gypsum-board dividing-walls which have a low penetration loss and relevant internal in-homogeneity \cite{BTD21}. 

Regarding weather and influence of rain in point-to-point mmWave links, the attenuation to long time exposure to rain in direct and side (NLOS) links at 25.84~GHz and 77.52~GHz have been measured and modelled in \cite{9232024}, showing a higher attenuation on the side links due to the scattering from the rain.


Human blockage also becomes more severe because of the directivity of the radio interfaces and the increased penetration losses with frequency. Frequently, two different approaches are used on the characterization of human blockage: an empirical, based on the analysis of fading statistics, and an analytical, based on the \gls{dked} model. The time variant human shadowing statistics at 60~GHz inside an Airbus 340 have been studied using different antenna types in \cite{5505046}. Similarly, short-range measurements at 60 GHz in \cite{6680667} show that the signal level attenuation follows a Gaussian distribution. In \cite{8254900}, 73.5 GHz measurements in pedestrian crowd scenarios have been used to derive a model based on Markov’s chain, which is also considered under the stochastic modelling approach in the 3GPP TR 38.901. On the other hand, analytical models based on the knife-edge diffraction principle predicts the obstruction loss based on the Huygens' principle: the diffracted front waves interfere constructively and destructively behind the blocker. 
Human blockage measurements in several mmWave bands up to 60~GHz have been compared to different modelling approaches in \cite{7986524, 8883197}, where a clear frequency dependence on the losses has been observed, with maximum losses up to 25~dB at 60~GHz. Similar results at 70~GHz are presented in \cite{7881087}, where measurements are compared to the predictions with the \gls{dked} model considering the directivity of the antennas. COST INTERACT contributions in human blockage at (sub-)THz have been analysed from measurements in \cite{ZKB23}. People with different complexion have been set frontally and laterally interrupting highly directional links, showing attenuation results higher than 25~dB and 30~dB, respectively. These results are underestimated with theoretical diffraction models as the METIS and DKED, which need to be extended including the antenna directivity, as discussed in \cite{7881087}.

Regarding the characterization of propagation at (sub-)THz, there are already several measurement campaigns in different scenarios, covering from very short link applications as desktop \cite{6898846}, rack‐to‐rack communications for server rooms \cite{8889517}; to middle-range applications as wireless links in meeting room \cite{9135404}, shopping mall and airport \cite{9448958}, between others; and outdoor applications as parking lot \cite{9473891}, courtyard and crossroad \cite{9810112}, train stations \cite{8684885}, etc. The impact of polarization in the channel when considering highly directive radio-interfaces at (sub-)THz has been studied in controlled experimental set-ups and in real complex environments in \cite{9135404}, where the dependence of the path gain on the incident and reflected angles is observed and contrasted with Fresnel. In recent times, (sub-)THz has also been targeted to industrial applications \cite{etsi22}, and therefore the characterization of propagation in this environments has gained a lot of relevance, \cite{9838910}. COST INTERACT contributions in (sub-)THz measurements in industrial settings and machines have been presented in \cite{DEV23}. The temporal/spatial characteristics of the channel in an external \gls{ap} to inside of machine (through penetration by protective glass) shows a channel with a rich set of multi-path components from the different metallic frames and machine components. A very precise idea of the location of the different objects in the environment can be depicted from the geometrical properties of these \glspl{mpc}. This offers an immense playground for the implementation of sensing applications that can be used to control different production processes. The polarization of the \glspl{mpc} also showed a behavior in concordance with the Fresnel equations. In addition, the obstruction of the LOS component by different parts of the machine or external items as a forklift truck passing by showed that there are remaining \glspl{mpc} that can be used to maintain a considerable link budget for communications. 

\begin{figure}[!t]
	\centering
	\includegraphics[width=\columnwidth]{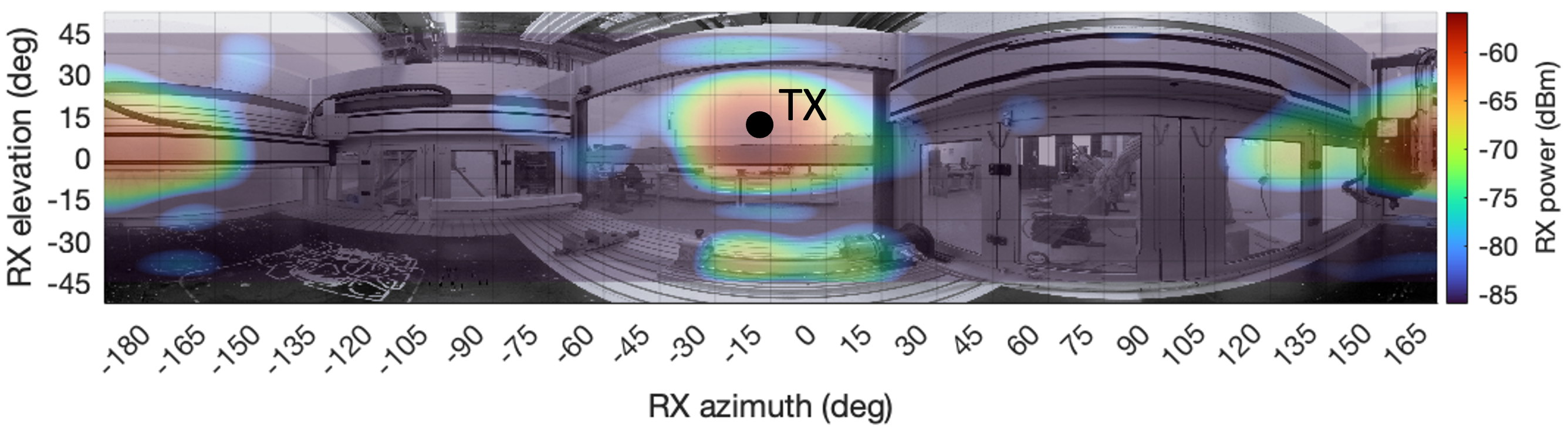}
	\caption{Power azimuth/elevation profile at RX in an external \gls{ap} to inside of machine scenario at 300~GHz, \cite{DEV23}.}
	\label{fig:5_2_fig1}
\end{figure}

However, despite the differences in propagation between the sub-6 GHz and higher bands, simultaneous multi-band measurements at different frequencies in several environments have yielded striking results in similarities: the dominant \glspl{mpc} are mostly present in the different bands \cite{8739860, 8887794, 9135630, 9768944}. 
Hence, there is a high correlation between bands, which has been further discussed and included in channel models~\cite{3gpp.38.901}. 
 
\begin{figure}[!t]
	\centering
	\includegraphics[width=\columnwidth]{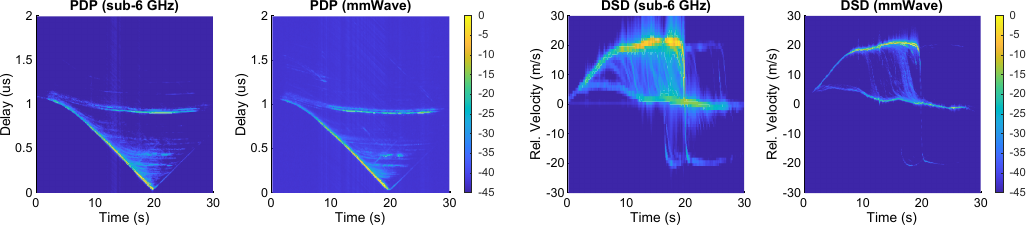}
	\caption{Power delay profiles and Doppler spectrum densities at 3.2~GHz and 34~GHz in V2V scenarios, \cite{10012965}.}
	\label{fig:5_2_fig2}
\end{figure}

COST INTERACT contributions in simultaneous multi-band measurements at sub-6~GHz, \gls{mmwave}, and (sub-)THz in different scenarios have shown strong similarities on the multi-path components in the different bands \cite{9768944, 10012965, RKA23}. The dominant paths are present with similar gain and, different with was expected, instead of an extremely sparse channel, dense-multi-path components have also been observed at \gls{mmwave} frequencies in many different scenarios. The \glspl{pdp} and Doppler spectral densities (DSDs) from multi-band (3.2~GHz and 34~GHz) V2V measurements in urban street scenarios showing strong similarities are displayed in Fig.~\ref{fig:5_2_fig2}, \cite{10012965}. This similar distribution of the \glspl{mpc} motivates to further explore the use of sub-6~GHz channel properties for beam-forming at \gls{mmwave}: the analysis of the multi-band measurements in an industrial scenario from \cite{9768944} shows that this can be effectively exploited in NLOS, \cite{DET23}. Similar results on the distribution of \glspl{mpc} in different bands have been observed in multi-band \gls{mmwave} and THz measurement in a conference room in \cite{RKA23}, where multiple common scatterers and the presence of high order reflections at THz have been observed.

Future measurements need to be designed to cope with the requirements on channel models to cover the spectrum from sub-6~GHz to THz and consider the needs of \gls{isac} applications. 
However, in case of the (sub-)THz bands, measurements are still rudimentary due to the complexity of the channel sounders and set‐ups required to capture the multi‐dimensional properties of the channel, making it often impossible to resolve  different dimensions simultaneously. Therefore, significant work needs to be spent on the joint estimation of Doppler, angular, and delay structure of the \glspl{mpc} at higher frequencies.

\newpage
\subsection{Measurement data collection contributed to COST INTERACT WG1}\label{sec:Measurements}

Empirical measurements are fundamental for the design, understanding and validation of radio channel models. Furthermore, the creation and sharing of standard reference datasets enables data-driven and hybrid model design approaches in an effective and reproducible way. COST INTERACT's open collaborative environment constitutes a unique opportunity to share measurements, simulation scenarios and models inside and outside the action. In this regard, the Horizontal Activity group 1 (HA1) is responsible for datasets' setup and maintenance. A total of four datasets, which are briefly reported and described in the following and made available at \url{https://interactca20120.org/wgs/datasets-2/}, have been collected by individual institutions and publicly shared to support the research activities and scientific collaboration within WG1. While we briefly describe the up-to-now contributed datasets below, WG1 remains open to further contributions, which will be made available at the same website. 

\begin{enumerate}

    \item \textbf{Indoor high-speed channel sounding at sub-6GHz and \gls{mmwave} \cite{Pasic2022}}

    Vienna University of Technology conducted measurements to compare sub-6GHz (2.55, 5.9 GHz) and \gls{mmwave} (25.5 GHz) indoor wireless channels in a high-speed scenario. For all measured scenarios, the wireless channel is measured with the same transmitter and receiver antenna positions, but with different center frequencies and velocities. This allows a direct comparison of the measured wireless channel in terms of fading environment and channel statistics. Results are provided in terms of time-variant channel transfer functions for discrete-time (snapshots) and frequency (subcarriers).

    \item \textbf{UPCT Indoor 5G measurements at 1-40 GHz \cite{electronics9101688}}\\
    UPCT conducted indoor \gls{los} \gls{mimo} measurements in the frequency range from 1 GHz to 40 GHz. Using a Vector Network Analyzer (VNA) setup, characteristic parameters including the relative received power, path loss, root mean square \gls{ds}, and K-factor were extracted and compared with ray-tracing simulations.

    \item \textbf{Measured dataset for performance analysis of wireless systems in real-world 60 GHz indoor channels \cite{Blumenstein2019IEEE, Liu2018Performance}}\\
    This dataset employs channel measurements of an indoor office environment at 60GHz from the measurement campaign held in Brno University of Technology (BUT), at the Department of Radio Electronics (DREL). The following measurements are employed \cite{Blumenstein2019IEEE} for the characterization of an indoor channel model for an IEEE 802.11ad single carrier physical layer (SC-PHY) MATLAB-based simulator, which is also provided.

    \item \textbf{Transmitter Identification and Fingerprinting based on RF Imperfections \cite{morin2019transmitter, Massouri2014CorteXlab}}\\
    These measurements are a compilation of the results \cite{morin2019transmitter} of experiments run on a series of datasets gathered in the Future \gls{iot}/Cognitive Radio testbed \cite{Massouri2014CorteXlab}. Such measurements can be employed for the detection of hardware imperfections in RF transmitters in order to identify a specific transmitter among others.

\end{enumerate}

\newpage
\section{Modeling methodologies}
\label{sec:Methodologies}
\begin{itemize}
    \item[] \textbf{by Tommaso Zugno, Enrico Vitucci, Nicola di Cicco, Diego Dupleich, Wei Fan, Ke Guan, Danping He, and Andrej Hrovat}
\end{itemize}

In the following sections, we present different channel modeling methodologies that have been proposed in the literature, including stochastic, map-based, ray-based, and \gls{ml}-based approaches. In each section, we describe the main concepts, overview the state of art, and summarize recent developments. 

\subsection{Stochastic models}
The characterization of wireless channels is of paramount importance for the design and evaluation of wireless systems, however, a deterministic knowledge of the channel behavior is difficult and impractical to obtain, as it requires to perform measurement campaigns or to run complex simulations. For this reason, a common approach is to adopt models which represent wireless channels as a stochastic process whose properties resemble real propagation phenomena.
In this regard, \gls{gscm} is one of the most widely adopted class of stochastic channel models. \glspl{gscm} adopt a stochastic approach to account for the presence of scatterers in the environment, and apply geometric properties to model signal propagation through multiple paths. The main advantage of this method is the possibility to easily represent different scattering environments by changing the parameters and/or probability distributions that model the physical properties of scatterers. 

\begin{figure}[h]
\begin{center}    
  \includegraphics[width=\columnwidth]{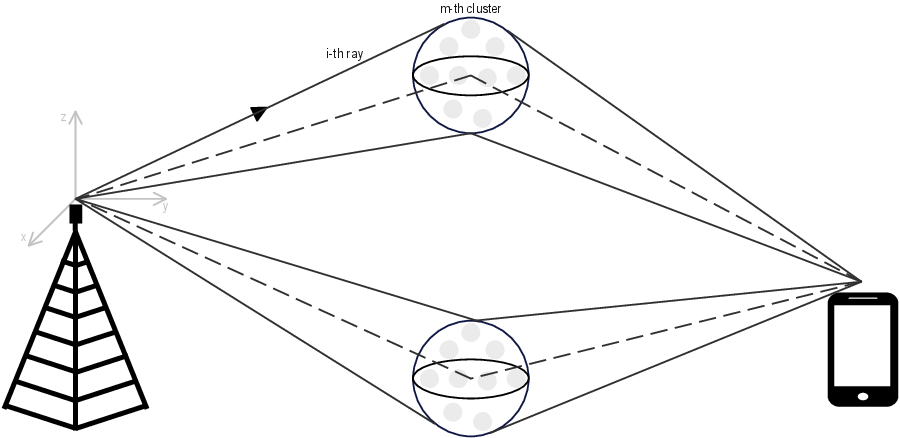}
\end{center}\vspace{-0.5cm}
\caption{Geometry-based stochastic channel model.}
\label{GSCM}  
\end{figure}

As represented in Fig.~\ref{GSCM}, signal propagation between a transmitting and a receiving node is modeled by the superposition of multi-path components (\glspl{mpc}), each representing a plane wave travelling along a different path. \glspl{mpc} depart from the transmitting node with different angles, and arrive at the receiving node from different directions, with different amplitudes, delays, and polarizations. 
Typically, \glspl{mpc} that exhibit similar characteristics are grouped together into clusters. Characteristics of each cluster (e.g., power, delay, AoA/AoD, polarization, etc.) are derived from random variables whose distribution depends on the scenario under consideration. 

\begin{figure}
\begin{center}    
  \includegraphics[width=\columnwidth]{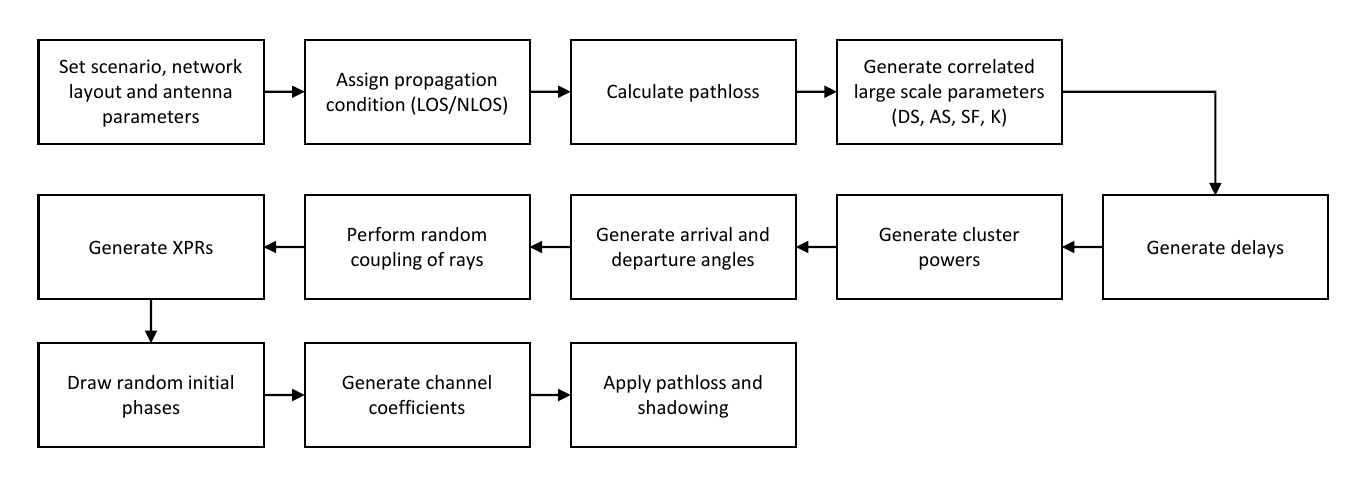}
\end{center}\vspace{-0.5cm}
\caption{Block diagram of the 3GPP \gls{gscm} model, from \cite{3gpp.38.901}.}
\label{38901}  
\end{figure}

Over the years, several works proposed \glspl{gscm} able to represent different scenarios and use cases, such as \gls{3gpp} TR~38.901 \cite{3gpp.38.901} (depicted in Fig.~\ref{38901}), COST 259 \cite{4027578}, Winner II \cite{winner2}, and Quadriga \cite{6758357}.
In particular, \gls{3gpp} TR~38.901 was selected as the reference model for the evaluation of cellular systems by the standardization community. This models supports a wide frequency, between 0.5 and 100 GHz, and the modeling of different propagation scenarios, including urban, rural, and indoor scenarios, within the same framework. Despite offering a high scalability and good generalization properties, it presents inherent limitations related to its fully-stochastic nature. For example, it does not provide a good support for the modeling of channel dynamics and inter-link correlation, thus preventing spatially-consistent evaluations. These issues can be overcome by adopting other approaches, such as the one described in \cite{6393523}, albeit at the cost of increased complexity and reduced number of applicable scenarios \cite{salous22}.

As part of the COST INTERACT action, \gls{gscm} models have been used for the performance evaluation of emerging use cases, such as vehicular and rail communications. In \cite{10008559}, authors used a \gls{gscm} to build a digital twin for assessing the reliability of vehicular communications, while in \cite{9723279}, authors used a \gls{gscm} to perform spatially-consistent real time simulation V2X scenarios. Moreover, in \cite{9343839} authors used a \gls{gscm} to train a frame error rate prediction algorithm for wireless communications among vehicles. 
In \cite{9769480} and \cite{dlr192826}, authors designed and validated a new model for train-to-train communications.

Other works proposed new approaches to overcome the limitations of current \gls{gscm} models for the simulation of next-generation wireless systems. Indeed, the introduction of novel paradigms, such as integrated sensing and communications, reflecting intelligent surfaces, and ultra-massive \gls{mimo}, require fundamental changes in the way the channel is modeled. For example, in \cite{9952200} authors extended the \gls{3gpp} TR 38.901 framework~\cite{3gpp.38.901} for the joint modeling of communication and sensing channels. They introduce the concept of \textit{sensing clusters} and describe the additional steps for the generation of the corresponding parameters. In addition to delays, powers, and angles, rays in sensing clusters are characterized by echo angles and radar cross section. Pathloss and shadowing are applied individually for each sensing cluster, and a multi-bounce model is applied to map scatterers to geometric positions.
In \cite{9625374}, authors proposed a \gls{gscm} for \gls{ris}-assisted communications which accounts for movements of terminals and clusters, and the time evolution of clusters in space. 
The channel impulse response is expressed as a summation of the direct path between BS and UT, and the indirect path reflected by the \gls{ris}. The reflecting properties of the \gls{ris} are modeled through the $\Phi$ matrix which includes the phase responses of the reflecting elements. A birth-death process is used to simulate the evolution of clusters in the space domain.
In \cite{9684731}, authors developed a \gls{3d} geometry-based double-spherical model for ultra-massive \gls{mimo} communications at \gls{thz} frequencies which accounts for the nano-material properties of plasmonic-based arrays.

\subsection{Map-based models} 
One of the issues regarding the \gls{gscm} approach is the modeling of spatial consistency among different links. To solve this issue, other approaches applying \gls{rt} principles based on simplified maps of the environment have been proposed \cite{7481518}. 

\begin{figure}[h]
\begin{center}    
  \includegraphics[width=\columnwidth]{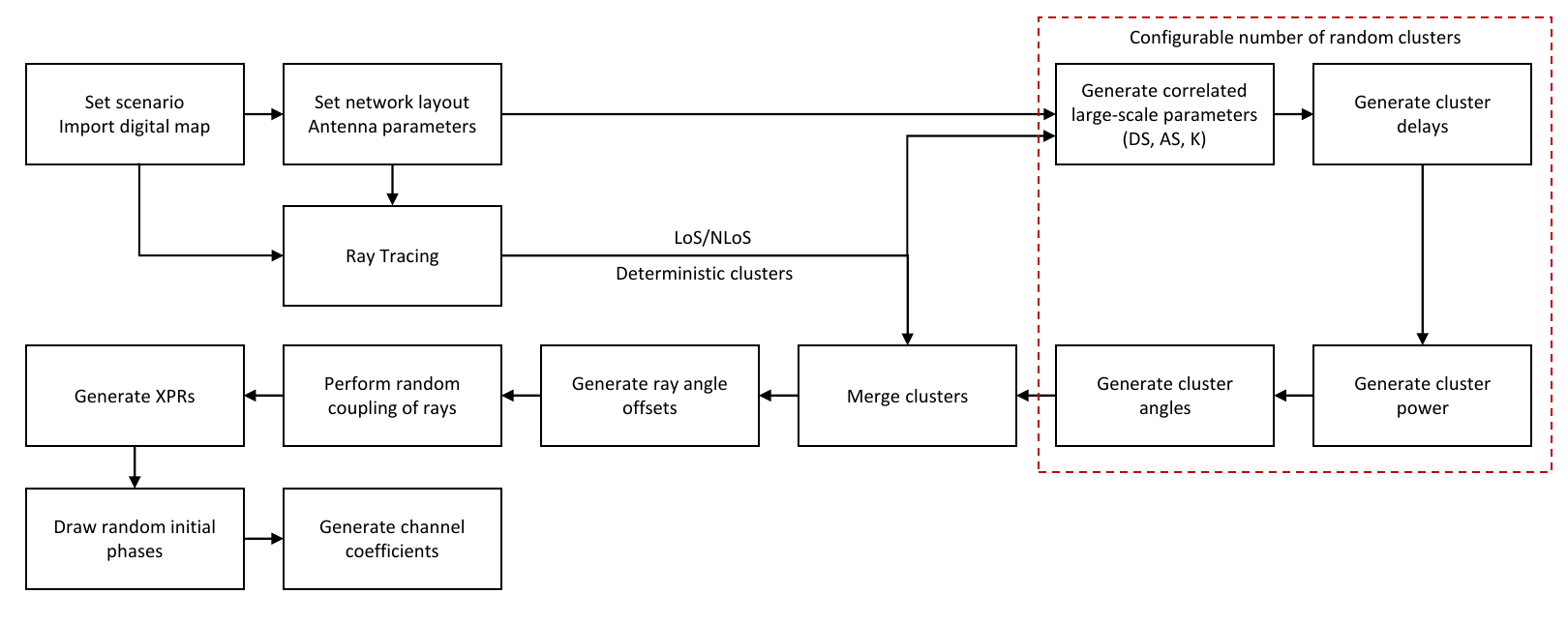}
\end{center}\vspace{-0.5cm}
\caption{Block diagram of the 3GPP map-based model, from \cite{3gpp.38.901}.}
\label{metis_proc}  
\end{figure}

An example of such approach is described in \cite{nurmela2015deliverable}. This model generates the channel response following the step procedure represented in Fig.~\ref{metis_proc}. The first step consists of creating the map, including the definition of transmitter and receiver positions, and the placement of blockers and scatterers. Then, a simplified \gls{rt} algorithm is used to compute the propagation paths and determine delays, departure and arrival angles. Finally, the channel impulse response is obtained by modeling the interaction of paths with blockers and scatterers according to the main propagation phenomena, including \gls{los} propagation, reflection, diffraction, scattering, shadowing, and antenna patterns.

Other models based on the same principle are available, the most popular being the \gls{3gpp} map-based model \cite{3gpp.38.901} and the NYUSIM model \cite{7501500}. 
More recently, the authors in \cite{9772928} proposed a map-based channel model for UAV communications at \glspl{mmwave}, while \cite{deng2022hybrid} presented a model for the evaluation of integrated sensing and communications. Moreover, a novel site-level deterministic model adopting a grid-based approach was presented as part of the COST INTERACT action \cite{9769688}. 

\subsection{Ray-based models}
\begin{figure}[h]
     \centering
     \begin{subfigure}[c]{0.49\columnwidth}
         \centering    
         \includegraphics[width=\columnwidth]{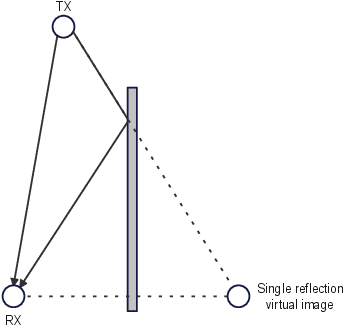}
        \caption{Ray tracing.}
        \label{RT}  
     \end{subfigure}
     \hfill
     \begin{subfigure}[c]{0.49\columnwidth}
         \centering    
         \includegraphics[width=\columnwidth]{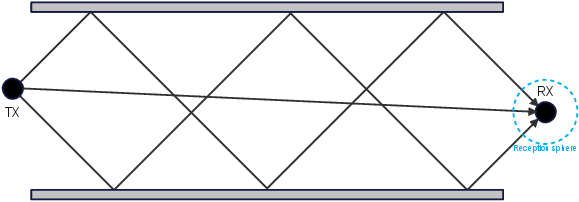}
        \caption{Ray launching.}
        \label{RL} 
     \end{subfigure}
\end{figure}

Ray-based approaches relies on the high frequency approximation, where electromagnetic waves are modeled as rays, following the principles of geometric optics for reflection and transmission. Two main approaches can be identified: (i) image ray tracing (\gls{rt}) and (ii) \gls{ral}. Ray tracing methods compute all rays that can reach the receiving point, e.g., by means of the Image Method represented in Fig.~\ref{RT}, then apply attenuation factors to each ray to account for the propagation phenomena. The main drawback of this approach is the computation time, which increases exponentially with the number of interactions.
In ray launching methods, multiple rays are launched from the transmitter into different directions according to a proper angular discretization, as depicted in Fig.~\ref{RL}. As in \gls{rt}, the field propagated by each ray is computed by taking into account all basic interaction mechanisms. Rays that reach the receiver contribute to the overall channel response, while those that miss the receiver or become too weak are dropped and not propagated further. In this case, the computation time increases with the number of launched rays; therefore, this parameter controls the trade-off between accuracy and computation time. 



When applied to highly dynamic scenarios, both methods still suffer from the problem of high computational complexity that prevents their use in real-time. One example is the case of vehicles moving at high speed and transmitting/receiving or generating scattering in a dense urban environment. Implementations on game engines like the one proposed in \cite{9723279}, solved this problem by adopting GPU parallelization. 
Other emerging paradigms like \gls{drt} can then be applied to mitigate complexity growth \cite{DRT}. Since the multipath structure remains essentially the same within a given time interval $T_C$, it is possible to predict the multipath evolution on the base of the current multipath geometry, assuming constant speeds and/or accelerations for moving objects within $T_C$, and using analytical extrapolation formulas. This is done without re-running a full \gls{rt} for every "snapshot" of the environment, therefore providing substantial savings in terms of computation time. The \gls{drt} is then a helpful tool for decreasing computational complexity and accelerating channel calculation, but also its use in real-time becomes possible. When \gls{drt} is embedded in a mobile radio system and used in real-time, ahead-of-time (or anticipative) channel prediction is possible, thus opening the way to interesting applications.
The \gls{drt} method is flexible enough to be employed either in a fully deterministic case, or when the path geometry is derived through statistical realizations of the environment, according to the \gls{gscm} approach. Therefore, a natural way of future development of highly dynamic channel modeling lies in further optimization of the execution time via merging the \gls{drt} approach with the game engine-based \gls{gscm} framework.

Another recent development is the integration of \gls{ris} models into \gls{rt}/ray launching algorithms in order to carry out realistic RF-coverage evaluations in \gls{ris}-enabled communication scenarios. In \cite{RIS-RayTracing}, a previously developed \gls{ris} macroscopic model based on a "Huygens-like" approach \cite{RIS-Reradiation} has been embedded in a \gls{rt} tool, and the performance of the \gls{ris}-based solution has been analyzed in simple reference scenarios by modifying a few simple parameters of the model. The results show that a gain of about 15–20 dB can be obtained in blind-spot locations with proper \gls{ris} placement and configuration, without the use of any additional active radio head, even when using simple designs such as pre-configured lossy phase-gradient metasurfaces. 

Future studies within the COST INTERACT action will deal with the development of a fully ray-based \gls{ris} model, which allows for better integration into RT algorithms and more realistic predictions involving multiple-bounce interactions where the \gls{ris} can be in any place in the interaction chain.

\subsubsection{Challenges for mmWave \& (sub)\gls{thz} channel modeling} 

For any \gls{rt} simulator, every object which exists within each scenario resembles kind of a Lego “building block”. Once all the building blocks are realized, any communication scenario can be implemented. With the available propagation models, the full-dimensional channel information can be obtained by employing the \gls{rt} methodology \cite{he2018design}. Nevertheless, the actual implementation of \gls{rt} techniques within the \gls{thz} band still faces significant challenges which need to be considered for future research \cite{yi2022ray}. Some of these challenges will be presented next as open research problems.

\textbf{Ultra-massive \gls{mimo} (UM-\gls{mimo})} systems generate very narrow beams to compensate for the very high path losses encountered in the \gls{thz} band. Since the \gls{rt} experiments are site-specific, the correlation of all the sub-channels created by UM-\gls{mimo} antennas can be characterized by \gls{rt}. However, with possibly several hundreds of antennas, the computational and storage capacities for such systems will dramatically increase \cite{spatial_channel_SNS_zhiqiang}. One possible solution to this problem is to develop a cloud-based \gls{rt} with high computation and storage capabilities.

\textbf{Antenna beam management} is another important challenge which needs to be carefully considered. Its main function is to steer the antenna towards the strongest ray/path and thus supporting the mobility of \gls{ue}. The current assumption for 5G \gls{mmwave} is that the antenna beam from the \glspl{bs} sweeps all possible directions every 5~ms, while the \glspl{ue} will transmit a short message as its response. However, since the \gls{thz} beam will become much narrower, the time required to check all possible directions will significantly increase. With the aid of \gls{rt}, firstly, the omnidirectional antenna can be used to directly identify the strongest path. Then, the \gls{bs} or/and \gls{ue} can select their own beam following the \gls{rt} simulation results in advance.

\textbf{Complex multipath} is another challenge caused by the “multi-structures” configuration. One solution for identifying the characteristics of the complex multi-paths is to integrate the individual transfer functions of the propagation graph with the aid of \gls{rt}. This is a new hybrid channel modeling approach which is based upon the join processing of \gls{rt} and graph theory. It is our belief that such hybrid approach is a promising approach for the accurate and efficient channel modeling of such dense \glspl{mpc}.

\subsection{ML-based approaches} 
Electromagnetic propagation is a complex phenomenon, as it depends on multiple, different factors, including the properties of the propagating signal (e.g., intensity, frequency, bandwidth, polarization, etc.), the geometrical and electromagnetic characteristics of the environment, and the specific position of the transmitter and the receiver inside the environment. As \gls{ml} methods are inherently fit to take care of complex problems \cite{AppliedML}, their use for wireless channel characterization has been attracting increasing attention over the last years \cite{huang2022artificial1}, \cite{huang2022artificial2}, \cite{MLoverview}.

If physical insight is heavily limited by the complexity of the target problem, a ML-based propagation model basically consists of a black box, that provides the existing, underlying pattern between some input data or between some output labels and the corresponding input features without any clear explanation. Conversely, \gls{ml} can be aimed at improving the accuracy of some baseline propagation model through the introduction of effective correction factors \cite{MLoverview} in case some physical/theoretical insight can be inferred.

To what extent a \gls{ml} propagation tool can reliably mimic the electromagnetic propagation process depends on the accuracy of the training stage, where a large amount of propagation data are effectively fed to the tool in order to learn the way the wireless channel actually behaves. Training therefore represents a crucial task that must be carefully planned and carried out. Propagation prediction through a well trained \gls{ml} tool is expected to be accurate and fast at the same time, to the extent that it can be relied on offline, i.e., for the design of wireless networks and systems, but also in real time, i.e., to assist the system (either end users or network equipment) during working operations.


We now provide a concise overview of the main learning paradigms in \gls{ml}, contextualized by some illustrative applications in propagation. We then elaborate on practical guidelines for choosing the most appropriate \gls{ml} algorithm for a given task, and we provide relevant pointers for future research directions of interest.
The fundamental learning paradigms in \gls{ml} are \gls{sl}, \gls{ul}, and \gls{rl}. In \gls{sl}, the task is to learn an input-output mapping given a finite dataset of input-output examples. \gls{sl} problems include, but are not limited to, regression, classification, forecasting, ranking, and segmentation. Examples of \gls{sl} problems in propagation are Path Loss regression \cite{thrane2020} and \gls{los} prediction \cite{dicicco2023}. In \gls{ul}, the task consists of learning patterns and/or extracting useful information from data. \gls{ul} problems may include clustering, dimensionality reduction, feature selection, density estimation, representation learning, and synthetic data generation. Finally, in \gls{rl}, the task consists of learning a policy that, after repeated interactions with a dynamic system (or environment), maximizes the long-term accumulation of reward signals. \gls{rl} is commonly applied to optimization and control tasks for which \gls{sl}, i.e., learning to imitate an optimal control policy, is unfeasible (e.g., due to the problem of deriving an optimal policy being intractable). Relevant examples in propagation include antenna tilt control \cite{vannella2022} and coordinated beamforming \cite{fozi2022}.

After identifying the learning paradigm that is most pertinent for a given task, the next crucial design point is the choice of the algorithm. In this regard, the “No Free Lunch” theorem states that all \gls{ml} algorithms are “equally bad” if averaged over all possible optimization tasks \cite{wolpert1997}. While this result may sound discomforting, it tells us that the choice of an \gls{ml} algorithm is mainly driven by the structure of the data at hand, i.e., one should choose an algorithm that can be proven (either theoretically or empirically) to be most efficient for the structure of the input data. Broadly, we can discriminate between tabular (or structured) data and unstructured data. As the name implies, tabular datasets consist of data that can be structured as a table, such that each row represents a single observation, and each column represents a distinct feature. For \gls{sl} problems, empirically, the best-performing models for tabular data are ensembles of \glspl{dt}, such as Random Forests and, more prominently, \gls{gbdt} models \cite{shwartzziv2022}. \gls{gbdt} models achieve a remarkable trade-off between ease of training, robustness to hyperparameters, model expressiveness and generalization capabilities, and are therefore advised when dealing with \gls{sl} problems on tabular data. Conversely, when dealing with unstructured data (such as images, graphs, point clouds, meshes, etc.), \gls{dl} model architectures become more prominent. Specifically, while the \gls{mlp} is a popular “one fits all” \gls{dl} architecture, it is advised that the choice of a \gls{dl} architecture is motivated by the presence of the appropriate “inductive biases” for the given data \cite{battaglia2018}. A relevant example are \gls{cnn}, which are among the state-of-the-art for Computer Vision problems \cite{Liu2022}. By learning local filters that are convolved with spatial 2D feature maps, \glspl{cnn} are able to efficiently exploit the spatial correlations in image data, i.e., the \gls{cnn} architecture possesses the right inductive bias for processing image data. Similarly, Recurrent Neural Network (RNN) architectures such as Long Short Term Memory networks (LSTM) \cite{hochreiter1997} and Gated Recurrent Units (GRU) \cite{chung2014} possess an inductive bias that makes them effective for processing sequence data. Note that an inductive bias, while desirable, is not mandatory for learning an effective model. In this regard, Transformer architectures \cite{vaswani2017} (the popular ChatGPT model \cite{brown2020} is one such example) have completely superseded RNNs in learning from large-scale natural language thanks to their expressiveness, scalability and ease of parallelization, albeit not possessing any particular inductive bias.

A recent family of models particularly relevant for propagation pertains to the field of Geometric Deep Learning \cite{bronstein2021}. Indeed, it can be argued that the laws of geometry are pervasive in propagation (e.g., in \gls{rt} algorithms and Stochastic Geometry), such that many fundamental algorithms in propagation operate on data that displays some geometric properties. In this regard, Geometric Deep Learning aims to devise model architectures able to exploit the underlying geometrical properties of the input data. The aforementioned \glspl{cnn} are a prominent example of Geometric Deep Learning applied to Euclidean domains (i.e., 2D feature maps). A generalization of \glspl{cnn} to non-Euclidean domains are Graph Neural Networks (GNNs) \cite{battaglia2018, bronstein2021}, which have been applied with success to physics simulation \cite{sanchez_gonzales2020} and processing of \gls{3d} point cloud data \cite{qi2017}. As such, Geometric Deep Learning holds an untapped potential for breakthrough applications in propagation.

Finally, one major limitation of complex \gls{ml} models is their lack of interpretability. Specifically, while tree-based models retain some degree of interpretability (e.g., by the means of feature importance), \gls{dl} models behave fundamentally as black-boxes, which may hinder their deployment in risk-sensitive application scenarios. With the goal of opening said black boxes and deriving precious insight on the learned knowledge, several eXplainable \gls{ai} (XAI) algorithms have been developed in literature. A prominent example are SHapley Additive exPlanations (SHAP) \cite{lundberg2017} which, given an arbitrary black-box function, computes the impact of every individual input feature to the final output. X\gls{ai} approaches for specific model architectures have also been developed. For instance, GradCAM \cite{Selvaraju2017} interprets \glspl{cnn} by highlighting on the input images where the model "looks" for taking a decision. GNNExplainer \cite{yin2019} interprets GNNs by deriving the subgraphs of the input graph that are the most influential for the output predictions. Recent advances in \gls{ml} interpretability, particularly relevant for propagation, are Symbolic Regression algorithms \cite{tenachi2023}. Briefly, the task in Symbolic Regression consists in finding the mathematical expression, modeled as a sequence of tokens (i.e., mathematical operators and physical quantities) that provides the best-fit to the data, balancing the delicate trade-off between goodness-of-fit and complexity of the derived expression. The associated optimization problem is of combinatorial nature: as such, \gls{rl} can be leveraged for efficiently exploring the space of all possible symbolic expressions. Overall, model interpretability can provide precious insight on the underlying physical laws present in raw measurement data, complementing domain-specific knowledge.

\newpage
\section{Channel parameter estimation}\label{sec:Estimation}

\begin{itemize}
    \item[] \textbf{by Wei Fan, Xuesong Cai, Diego Dupleich, Ruisi He, and Bo Ai}
\end{itemize}

\subsection{Parameter estimation techniques}
\textcolor{black}{Channel parameter estimation aims at extracting propagation parameters such as propagation delay, angles, polarimetric amplitudes, etc. of path components from the measurement data. COST INTERACT has contributed to the development of such algorithms. For example, the authors in \cite{9909796} proposed a maximum-likelihood estimation algorithm to deal with channels that exhibit a mixture of independent \glspl{dmc}, which is in contrast to the commonly assumed model of single \gls{dmc} mode. An auto-encoder was proposed to infer the order of \gls{dmc} modes and for initializing the parameters of the \gls{dmc}s.} In general, channel parameter estimation techniques can be mainly classified into four categories: spectra-based techniques, subspace-based techniques, sparsity-recovery-based techniques, and maximum-likelihood-based techniques.
\subsubsection{Spectra-based techniques}


Bartlett beamforming is the most classical spectra-based method, \textcolor{black}{aiming to find the direction(s) with dominant power(s) \cite{526899}}. Its variant, a frequency-invariant beamformer for uniform circular arrays can also be found in \cite{7523340}. \textcolor{black}{The basic idea was to pre-compensate to frequency-dependent phase variation so that \gls{2d}-\gls{fft}, i.e., beamforming,} can be efficiently applied to finding the dominant paths in azimuth and delay domains jointly. For \gls{mmwave} frequency bands, a widely applied spectra-based method is to obtain the joint angle-delay spectrum according to the channel measurements by rotating horn antennas to different directions \cite{8103059,8094309,7400962,7109864}. These methods are straightforward and with relatively lower complexity. However, system responses such as antenna radiation patterns are usually embedded in the resulting spectra, making it difficult to separate the propagation channels from the sounder hardware, not to mention the low resolutions. 
\subsubsection{Subspace-based techniques}
Among the subspace-based methods, the well-known ones are \gls{music} \textcolor{black}{\cite{1143830}} and \gls{esprit} \textcolor{black}{\cite{1457851}}, and their variants can be found in, e.g., \cite{8093607,1395953,575559,8316940,552142}. In the early development of these techniques, the basic assumption is that propagation paths are uncorrelated so that the covariance matrix of the received signals can be decomposed into signal subspace and noise subspace \textcolor{black}{\cite{1143830}. By examining the distances (orthogonality) between steering vectors and the noise space, a spectrum can be obtained with its peaks indicating path directions. Intuitively, if a steering vector is orthogonal to the noise space, it means that it belongs to the signal space, i.e., contributing to the  received signals.} Later on, these techniques were extended to dual domains such as delay vs. frequency covariance matrix. The main limitation is the deficiency in resolving a large number of paths and requiring a certain number of snapshots to obtain a sample covariance matrix. 
\subsubsection{Sparsity-recovery-based techniques}
Sparsity recovery algorithms \cite{7390019,8122055} are developed based on the assumption that the channel exhibits sparsity in parameter domains, \textcolor{black}{i.e., only a few paths contributing to the received signals}, although the assumption is still questionable even in \gls{mmwave} frequency bands \cite{8386686}. By exploiting specific optimization principles, e.g., convex optimization, channel parameters can be recovered. 
\subsubsection{Maximum-likelihood-based techniques}
Despite the high complexity of maximum-likelihood-based estimation algorithms, they can extract the propagation parameters that are properly defined in the signal model with a high resolution that approaches the theoretic lower error bound, i.e., \gls{crlb}. The most widely used maximum-likelihood-based algorithm is the \gls{sage} algorithm \cite{753729}. It is a further enhanced algorithm based on the \gls{em} principle \cite{543975} that is theoretically proven to converge to a local maximum of the likelihood objective function. In \gls{sage}, the high-dimensional estimation/optimization problem can be decomposed into several one-dimensional problems, leading to much lower complexity and faster convergence. In \cite{8713575,9104014,9115069}, variants of the \gls{sage} algorithm can also be found for \gls{mmwave} wideband large-scale arrays, where the trajectories in delay domain across the array aperture are exploited for effective initializations and interference cancellation. The \gls{sage} algorithm usually assumes that the propagation paths are well-resolvable. However, it is possible that due to the scattering effect of rough surfaces, the resulting \gls{mpc}s can be very dense in delay and angle domains that cannot be well resolved by the intrinsic ability of the sounder. In such cases, one needs to modify the signal model to consider these dense \gls{mpc}s, i.e., \glspl{dmc}, as colored noises (in addition to white Gaussian noises) with certain power profiles for a better estimation of other discrete \gls{mpc}s, which is basically the RIMAX algorithm \cite{RICHTER1}. Although there exist not a few different parameter estimation techniques, future \gls{mmwave} and \gls{thz} propagation channels will pose more challenges due to much larger bandwidths, much larger array apertures, etc. Spherical-wave propagation, channel birth-death on the array, frequency-dependent responses of antennas and sounders, etc., will make the signal model much more complicated, meaning that low-complexity yet still high-resolution estimation techniques are still in need.

Another important consideration for parameter estimation is that the system response of the channel sounder must be well characterized for a realistic signal model. Otherwise, the mismatch of the signal model from reality can result in many ghost (erroneous) paths estimated \cite{THzMagazine}. The response of cables, power amplifiers, filters, converters, etc., of the channel sounder can be easily calibrated through back-to-back measurements, i.e., directly connecting Tx and Rx without antennas. For antennas or antenna arrays, they can be placed in an anechoic chamber to measure the \gls{3d} responses of antenna elements at discrete angles. The measurement data can be later on exploited to interpolate the responses at arbitrary angles. There are different ways to do this, which include direct (linear or non-linear) interpolation, spherical harmonics, and \gls{eadf} \cite{EADF}. Direct interpolation is straightforward but non-analytic. Alternatively, one can transform the measured pattern from the spatial domain to another domain using basis functions, either spherical harmonics or Fourier basis (\gls{eadf}). Using a forward transform, the transformed spectra are obtained and then utilized to recover spatial patterns. There is also a possibility to compress the measurement data if the spectra are concentrated so that the unimportant components in the transformed domain can be removed. A practical application of \gls{eadf} in dynamic \gls{mmwave} channel sounding using 128$\times$256 switched arrays can be found in \cite{mmWavesounder}.

\subsubsection{Future challenges and directions}
6G is envisioned to support applications beyond the current
5G mobile use scenarios, which will pose stringent
requirement on the communication systems in terms of data-rate, latency, reliability, and so on. To meet those demanding requirements, a raft of advanced radio technologies are envisioned, e.g.  utilization of  higher
frequency spectra (e.g. up to sub-\gls{thz} frequency bands), higher system bandwidth (e.g. up to a few GHz), and larger-scale
multi-antenna systems (e.g. large-scale or extremely large-scale antenna), utilization of \gls{ris}, etc. The advancement in 6G radio technologies have posed significant challenges to the channel estimation parameter. For example, the extremely large scale antenna array will bring a few new challenges for the channel parameter estimation: near-field and spatial non-stationary effects. State-of-the-art algorithms, which are developed based on plane
wave model and spatial stationary channel might fail to address all the new features introduced by the massive \gls{mimo} systems. The narrow-band assumption might also be violated for 6G radio systems that potentially utilize the ultra-wide-band technology, which should be properly considered for developing channel parameter estimator. As for the sub-\gls{thz}/\gls{mmwave} channel measurements, the required phase accuracy might be difficult to achieve, though a few works on phase compensation concept have been reported to achieve coherent phase measurement for virtual antenna array systems. Therefore, it would be desirable to develop channel parameter estimators, which is robust to phase measurement inaccuracy. 

\subsection{\gls{mpc} Clustering} 
Multipath clustering and identification have been important for channel modeling. \gls{ml} naturally meets the need to group multipath components (\glspl{mpc}) with similar channel characteristics, and some well-designed clustering algorithms which naturally incorporate propagation characteristics have drawn great attention~\cite{he2018clustering,huang2022artificial1} for cluster-based channel modeling, and many researches have been conducted in COST INTERACT. Clustering algorithms can be generally categorized into: i) Shape-based; ii) Distance-based; iii) Density-based; and iv) Computer vision-based clustering.

\subsubsection{Shape-based clustering}
Shape-based clustering involves revealing cluster structure of \glspl{mpc} in power-delay domain. For example, \glspl{mpc} belong to the same cluster are supposed to have a PDP that follows a single exponential decay, according to the Saleh-Valenzuela channel model. To recognize clusters in PDP, a fitting method is used to check if envelope shape of \glspl{mpc} matches a particular distribution, as shown in Fig.~\ref{Clustering_figure1}(a). This method adjusts cluster members by seeking the best fitting result and it is able to accurately identify \gls{mpc} cluster~\cite{he2016clustering}. Several shape-based methods have been proposed: i) training a hidden Markov model (HMM) to learn distribution of \glspl{mpc} in PDP and optimizing cluster members using the Viterbi algorithm; ii) using an observation window to separate large cluster and applying a threshold of slope to improve clustering accuracy on small clusters; iii) using kurtosis to measure peakedness of a distribution and applies region competition to divide \glspl{mpc} into different clusters~\cite{gentile2013using}. In summary, shape-based cluster identification has several advantages including not requiring much prior knowledge about number of clusters and having a relatively low computation complexity. However, its limitation is the lack of angle information during clustering, which can impact identification accuracy.

\subsubsection{Distance-based clustering} 
\begin{figure}[!t]
	\centering
	\includegraphics[width=\columnwidth]{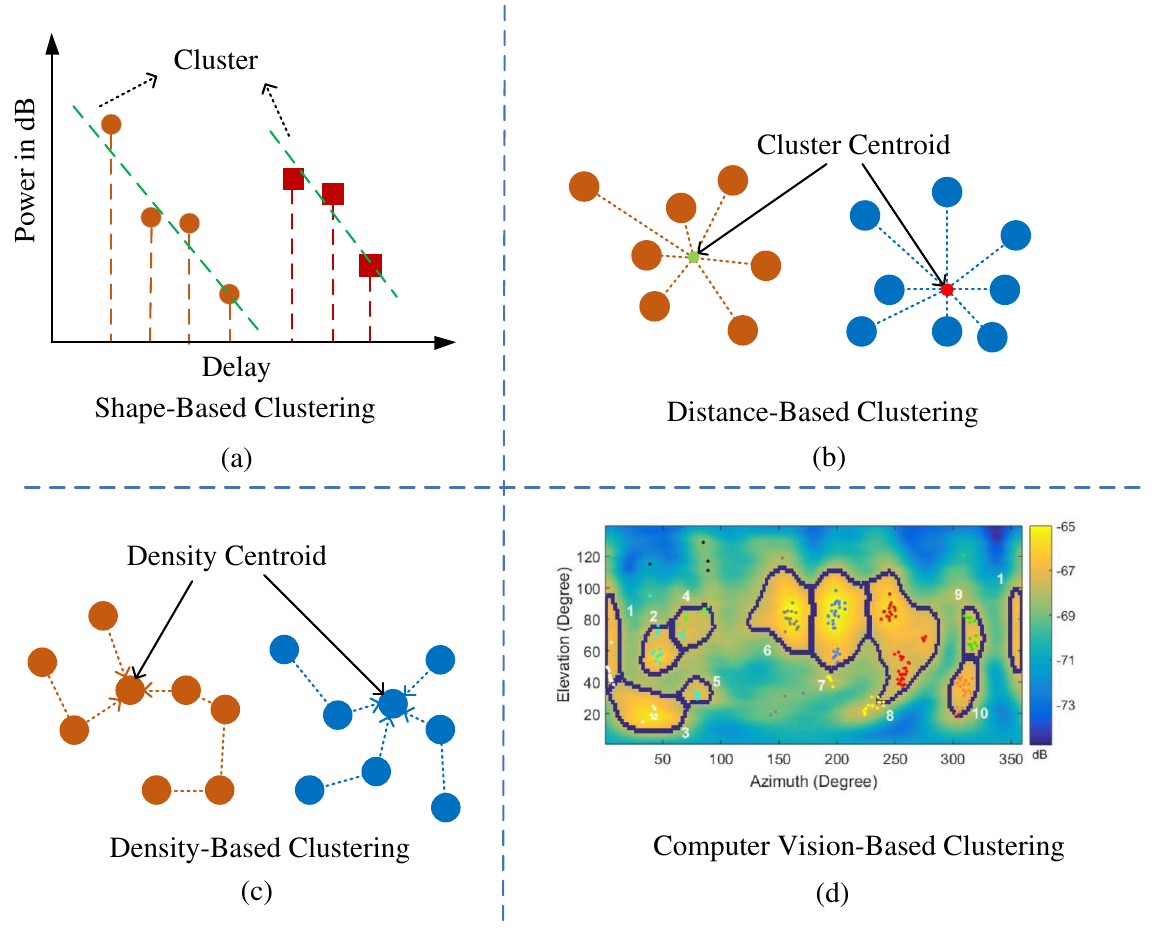}
	\caption{Illustration of different clustering algorithms.}
	\label{Clustering_figure1}
\end{figure}

Distance-based clustering measures similarity between different \glspl{mpc} based on distance , which is defined in terms of delay, angle of arrival (AoA), and angle of departure (AoD), as shown in Fig.~\ref{Clustering_figure1}(b). Commonly used distance measures for \gls{mpc} clustering include squared Euclidean distance (SED), normalized Euclidean distance (NED), and multipath components distance (MCD). SED focuses on natural difference between each parameter, while NED focuses on ratio difference. Sequential clustering-based algorithms have been proposed to identify \gls{mpc} clusters using SED. However, to compare parameters in different domains, it is necessary to normalize parameters to the same scale. MCD has been proposed to address this issue by normalizing delay and angle before calculating distance, which is found to have fairly good performance~\cite{steinbauer2002quantify}. Further, many distance-based clustering algorithms such as hierarchical tree clustering, K-power-means (KPM), and fuzzy-C-means (FCM) have been proposed and widely used.

\subsubsection{Density-based clustering} 
In the natural propagation environment, \gls{mpc} are grouped and the \gls{mpc} near cluster centroid usually have higher density than those at the edge of cluster. Density-based clustering algorithms, such as DBSCAN, can identify \gls{mpc} clusters based on distribution and density property, as shown in Fig.~\ref{Clustering_figure1}(c). These algorithms calculate \gls{mpc} density for clustering and do not require prior knowledge such as cluster number and cluster centroids initial position. However, how to measure \gls{mpc} density is important to algorithm performance. To obtain improved performance,~\cite{he2017kernel} firstly proposes the perspective to consider physical propagation mechanism for clustering algorithm by designing \gls{mpc} density. It proposes a novel aspect of \gls{mpc} kernel-power-density to well incorporate the modeled propagation behavior of \glspl{mpc} into clustering algorithm, and it obtains significant performance gain in terms of low computational complexity and high clustering accuracy.

\subsubsection{Computer vision-based clustering}
Computer vision-based clustering exploits image processing methods to identify \gls{mpc} clusters based on visual criteria such as shape of potential cluster, distribution pattern of \gls{mpc} delay and angle, as shown in Fig.~\ref{Clustering_figure1}(d). One example is the Hough transform-based clustering, which uses the Hough transform to recognize trajectory of \glspl{mpc} in delay domain and merges the recognized trajectory into clusters. Another example is the PAS-based clustering and tracking (PASCT) algorithm~\cite{huang2018power}, which uses the maximum-between-class-variance method to separate potential cluster groups from background noise and further divides the clusters by using density-peak-search method. These methods follow an intuitive approach and provide results that conform to human observation whereas also benefit from the rapid development of computer vision.

\subsubsection{Future Work}
Many long-standing problems remain unsolved on this topic. For example, for time-varying non-stationary channels, the existing algorithms still need to be improved significantly. Supervised AI-based clustering and tracking methods are worth receiving more attention in the future, especially considering the phenomenal increase in both the amount of collected channel measurement data and the available computing power. Moreover, drawbacks still exist with respect to algorithm complexity, threshold choices, and assumptions about prior knowledge. Algorithms with few prior knowledge about clusters should be further developed.

\newpage
\section{New technologies} 
\label{sec:NewTechnologies}

\begin{itemize}
    
 \item[]\textbf{by Joonas Kokkoniemi, Marco di Renzo,  Narcis Cardona, Ruisi He, Bo Ai, Wei Fan,  Xiping Wang, Ke Guan, Tomaz Javornik, Franco Fuschini,  Yang Miao, Mi Yang, Dan Fei, Guido Valerio, and Julien Sarrazin}

\end{itemize}

\subsection{Path Loss Measurements and Modeling for \gls{ris}} 
\label{sec:RIS}


In recent years, \gls{ris}s have been under active discussion and investigation as a promising technology to enable future 6G networks. \gls{ris}s are artificial electromagnetic surfaces comprised of large numbers of sub-wavelength unit cells or antenna elements. Those are either very small elements ($\ll\lambda$) forming a larger artificial surface (a metasurface) or individually controllable antenna elements such as in reflect arrays. The \gls{ris}s can flexibly control the parameters of wireless signals, such as phase, amplitude, and polarization, thus enabling the emerging concept known as “smart radio environment” \cite{DiRenzo2020}. The core principle of the RISs is to be able to capture a part of the radiated power with the aperture of the RIS and redirect it via reflection towards wanted direction by manipulating the phase (and amplitide) of the insident radiation. Usually the goal is to do beamsteering the amplify certain directions, but as mentioned above, it is also possible to manipulate the signal itself. Therefore, the RISs allow the manipulation of the radio channels. This is traditionally impossible and usually the only way to manipulate the signals is via antennas and arrays of those. Hence, the RISs give an interesting opportunity for added control that can be used to overcome problems, such as signal blockages and cell edge signal amplification.

Two important use cases for RISs in the literature are shoot-through RIS and reflective RIS. The former can be used for beamforming close to antenna, but the latter is more important what comes to channel modeling and especially manipulation of the channel coefficients. In this section, we talk exclusively about reflective RISs and the channel modeling related to those. So far, some of the challenges with RIS channel models is the rather limited number of physical prototypes and multitude of options and ways to manufacture RISs. Later in this section we give some examples of the channel modeling activities related RISs.

\subsubsection{RIS channel modeling challenges}

Generally speaking, some of the challenges in RIS channel modeling arise from the fact that the behavior of the RIS itself depends on how it's made and how it as been configured. The basic communication scenario is that we have a Tx, RIS, and an Rx. The channel between Tx and RIS and RIS and Rx are traditional channels, e.g., free space of fading channels. The phase shifts introduced by the RIS are then optimized minimize the total loss in the cascaded Tx-RIS-Rx channel. Because the RIS is an active element, often assumed to be controlled by the base station, a closed formed macroscopic channel models are hard to derive as the RIS by definition reconfigures the channel coefficient based on the particular situation. Therefore, also the total path loss depends on the entire system setting and outcome of the optimization problem.

Whereas wireless communication engineers in the past have resorted to macroscopic channel models in system analysis and optimization, there are also other challenges what comes cascaded RIS channels. For instance, the real systems are almost never random. The network engineers design the network based on the maximization of the signal power in some local are and availability of locations for the base station and supporting hardware. In the case of RISs this can mean that the engineer may try to arrange a very good LOS channel between the base station and RIS whereas the RIS-user channel can be normal mobile fading channel, for instance. But this is just one of many options from network design point of view and the individual channels depend on the location and how the network elements can be placed. This makes the general channel modeling very challenging and the channel losses and potential gains achieved by RISs depend on the deployment environment. Especially in rural setting where distances are long, RISs are most likely not going to provide good gain as the reflected energy decreases fast with distance. However, in the case of urban and indoor scenarios where RISs are expected to be the most beneficial, there are powerful tools avaialable to test and optimize networks via simulations.

Ray tracing has become an important tool to study channel especially at high frequencies where traditional channel modeling is challenging due to difficult channel measurements. As the ray tracing relies on fixed 3D maps, they have also been used extensively in network design in order to optimize the network element placement. Ray tracing is therefore also a very good tool in studying RIS channels in various scenarios, e.g., as shown in \cite{Pyhtila2023} in urban environment. The downside of the ray tracing is that the simulation always require accurate 3D models of the desired environment. The upside is that testing the network elements in simulation environment is very flexible and fast. Therefore, the ray tracing is very good tool to test and evaluate the RISs as well. This is somewhat related to digital twins in which we have a replicated digital environment where we can test, e.g., RISs, but theoretically also optimize those in real time in the case of actual physical deployment.

Ray tracing is particularly powerful in system evaluations and performance testing, but can be locally used for channel modeling as well. The challenge in channel modeling tends to be that the results are representative for that particular environment, but not in general. However, it is still much faster to generate data with ray tracers that by real measurements. Therefore, an appealing option is to calibrate ray tracers with real measurement data in order to extend the measured data. In the case of RIS this still does not take away the difficulty of statistical modeling of an active reconfigurable element that can take many forms and sizes. Ray tracing can still be very efficiently used in evaluation of the RISs in the desired environment in order to deside where to place it/them and what kind of gain do those give in the scenario in hand. Whereas there are too many channel scenarios to discuss herein where RISs could be useful and what those require from the channel modeling, in below we give some examples of the works related to RIS channel modeling.

\subsubsection{RIS channel models from literature}

The path loss and propagation are basic characteristics of wireless channels. The channel represents the basic relation between the wireless signal power and the transmission distance, which can provide information on how far a wireless signal can be successfully transmitted. Due to the importance of characterizing the path loss and channel reciprocity characteristics of \gls{ris}-assisted wireless communications, researchers have recently conducted related studies that are based on different analytical assumptions and approaches \cite{Tang2021a},\cite{Tang2022}.

Di Renzo et al. \cite{DiRenzo2020a} derived asymptotic scaling laws of the path loss as a function of the transmission distances and the size of the \gls{ris} in the far-field and near-field cases. The results are obtained by leveraging the scalar Huygens-Fresnel principle in a two-dimensional space. Garcia et al. \cite{Garcia2020} calculated the radiation density of the scattered field in the near-field and far-field under the assumption of dipole antennas and discussed the scaling laws as a function of the transmission distances numerically. Ellingson \cite{Ellingson2021} proposed a physical model for the path loss of an \gls{ris}-assisted wireless link under the assumption that the antenna gain of the transmitter/receiver is constant over the \gls{ris}. Najafi et al. \cite{Najafi2021} developed a physics-based \gls{ris} path loss model, in which the impact of grouped unit cells on the wireless channel is obtained by solving the integral equations for electromagnetic vector fields under the far-field assumption. Danufane et al. \cite{Danufane2021} generalized the model in \cite{DiRenzo2020a} to a three-dimensional space by using the vector generalization of Green’s theorem, and characterized the scaling laws of the path loss as a function of the transmission distances and the size of the \gls{ris} based on scattering theory. Wang et al. \cite{Wang2021} proposed a radar cross section-based path loss model that introduces an angle-dependent reflection phase behavior of \gls{ris} unit cells. Gradoni and Di Renzo \cite{Gradoni2021} developed a path loss model that is based on the theory of mutual impedances of thin dipole antennas. The end-to-end channel model resembles a \gls{mimo} communication system and considers the mutual coupling among \gls{ris} unit cells. Recently, Di Renzo et al. \cite{DiRenzo2022} proposed a path loss model, based on scattering theory, assuming that the incident signals are not constant over the \gls{ris} elements. A summary of the above-mentioned research works on \gls{ris} path loss modeling is available in Table I of \cite{Degli-Esposti2022}.

\subsection{Channel Measurements and Modeling for \gls{isac}} 

In the last decade, radar has been considered in some works as the natural complement to communications, 
in which the radar-based sensing is a default technology. 
In \cite{dokhanchi2019mmwave} a low-complexity algorithm for joint radar and communications in automotive is developed, while in \cite{kumari2017ieee} the sensing from radar is used to assist the beam-steering at mmWave links. The sensing information is used to predict the communications channel in \cite{ni2021uplink}.
Nevertheless, when we refer to \gls{isac} in
this paper, it is done for communications-centric \gls{isac}, meaning that sensing is implemented using the communication signals from one or more radio access nodes. 
To this end, in COST INTERACT, authors in 
\cite{rinaldi2021dual} are developing solutions for \gls{jsac}, with parallel sensing and communication waveforms that share the same bandwidth.

\gls{isac} has become a design paradigm for 6G, for which understanding and modeling the behaviour of the sensing and the communications channels simultaneously is crucial. The main \gls{isac} applications in 6G mobile networks are vehicular communications (V2X), sensing as a service, remote sensing and environmental monitoring, while those applying to short distance wireless channels are, among others, in-cabin sensing, smart home, and human-computer interaction \cite{tan2021integrated, han2022thz, cui2021integrating}. This confers different \gls{isac} channel scenarios, which can be summarized into three types (Fig.~\ref{fig:isac-scenarios}): monostatic, bistatic, and distributed, depending on the deployment of the transmitter and receiver(s).



\subsubsection{Sensing scenarios: monostatic, bistatic and distributed}

\begin{figure}[h]
  \centering
  \includegraphics[width=0.95\textwidth]{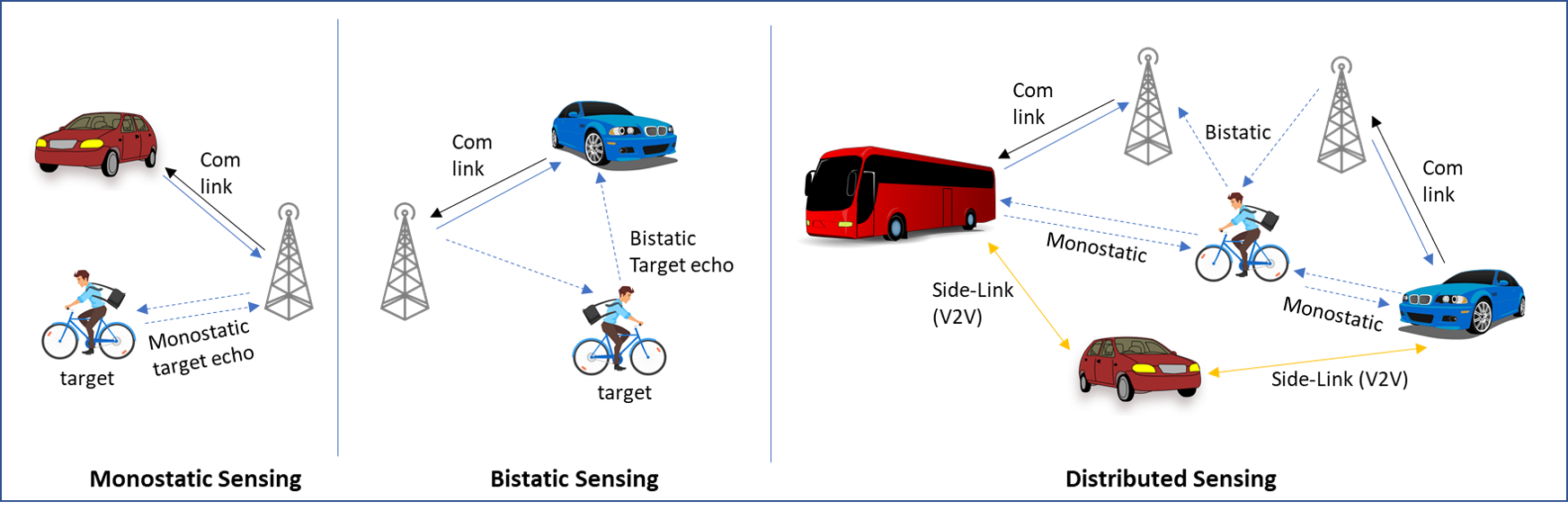}
  \caption{Monostatic, Bistatic and Distributed \gls{isac} scenarios}
  \label{fig:isac-scenarios}
\end{figure}

\gls{isac} scenarios are monostatic when sensing transmitter and receiver are at the same position; bistatic case refers to  sensing transmitter and receiver being separated, while distributed scenarios account for multiple sensing paths available for the same target. In~\cite{thoma2022characterization} a description of the monostatic, bistatic and
distributed \gls{isac} concepts, and the advantages of the centralized version of \gls{isac},
are discussed. Figure~\ref{fig:isac-scenarios} summarises the three cases.

For the monostatic case, many radar concepts apply, like the radar cross section (RCS) to measure the radar reflectivity of individual targets. Monostatic sensing channels can anyway be modeled from the multipath propagation, limited to the case of having the transmitter and receiver at the same position, so most of, but not all, the sensing energy coming from \gls{los} components. 
To that end,  \cite{li2022rough} provides an empirical scattering model obtained from measurements of materials at 28 and 140~GHz. The implementation of monostatic \gls{isac} sensing requires the full-duplex mode of the radio interface, which is not implemented so far in current 3GPP standards. This is the reason why some proposed systems for \gls{isac} refer in fact to a \gls{jsac} strategy, in which a radar sensor is used to improve communication links.

The estimation of the communications channel from the sensing one is investigated in \cite{li2022novel}, where the authors consider the communication receivers to be the targets of a monostatic OTFS-based \gls{isac} system. Thus, the two channels in this case are highly correlated.
In \cite{gaudio2020effectiveness} the authors use OTFS modulation for joint radar parameter estimation and communications, exploiting the channel sparsity in the Delay-Doppler domain for efficiently separate sensing (radar) parameters estimation and communications.

In the bistatic case, 
the transmitter sends communication and sensing signals and the receiver captures the echoes from targets and clutter together with the multipath (scattered) components that compose the communication channel. As a consequence, any model that describes simultaneously the sensing and the communications channels has to be geometrically accurate for simultaneous multiple links, and reproduce moving targets along different tracks with spatial consistency, including phase continuity \cite{thoma2022characterization}.



\subsubsection{Trends in \gls{isac} channel modeling}
The \gls{isac} channel models are a combination of two channels in one single geometry, i.e., a common framework that develops two approaches, either simulated with more or less precision, or stochastically created. 
In \cite{cardona2023integrating}, the \gls{isac} channel is defined as a two-port system in which sensing and communication channels, as well as the common interference between them, are modeled as the reflectivity and transmission functions of such system. This scheme is adaptable to any current model in the literature, either deterministically created by precise 3D ray-tracing, stochastically from a random distribution of scatterers, targets, and clutter, or a hybrid of the two.

\paragraph{Stochastic models}
Recent approaches on \gls{isac} 
stochastic channel modeling propose \gls{gscm} extensions to the 3GPP channel model~\cite{3gpp.38.901}, e.g., within COST INTERACT~ \cite{lopez2022considering18} and elsewhere~\cite{zhang2022general}, \cite{liu2022shared17}. The stochastic channel modeling methods dominated the evaluation of wireless communications in 5G due to their low computational complexity and easy standardization, which was sufficient to evaluate the communication performance of 5G use cases. 
The application of \gls{gscm} models to \gls{isac} requires adding some important elements to the model, since communication and sensing channels are somewhat different. For example, \cite{liu2022shared17} includes both stochastic and deterministic approaches while accounting for spatial consistency. 
In~\cite{lopez2022considering18}, paths are generated by probabilistic functions derived from channel measurements made in real scenarios, thus geometrically pre-setting the distribution of effective scatterers, i.e., the objects on which the set of rays is incident. The authors in \cite{liu2022shared17} propose a shared cluster-based stochastic \gls{jsac} channel model and conducts a channel measurement campaign in typical \gls{los} and N\gls{los} indoor scenarios at 28~GHz and obtains the power angular delay profiles (PADPs) of the communication and sensing channels. 
For monostatic case, the correlation between the channels may exist, but this is not so obvious correlation for bistatic and distributed cases, mainly because sensing ``targets'' and communication ``scatterers'' are not necessary the same objects or, if they are, do not have the same reflectivity behaviour.


\paragraph{Deterministic models}

Current deterministic channel models have modeled scatterers for communications and 
have considered the reflection from them in the form of specular reflection or diffuse scattering. From radar/sensing perspective, any target in the scene can be modeled according to its RCS, but for \gls{isac}, more detailed parameter are required, e.g., the bistatic target reflectivity (BTR) \cite{thoma2022characterization}. 
An example of a target is a person on a bike (Fig.~\ref{fig:isac-scenarios}), a moving target with local movements, causing in terms of sensing a long-term Doppler because of its displacement, and micro-Doppler originated by the person and wheels movements \cite{thoma2022characterization}. The same work further establishes that, for \gls{isac}, deterministic models with more physical descriptors are needed, especially for target modeling, in contrast to the statistical approach of the \gls{gscm} models. To resolve the issue, the authors propose a propagation channel model for \gls{isac} that is geometrically correct for multiple simultaneous sensor links and reproduce a moving target in a spatially consistent way along a track, which includes phase continuity. 

\paragraph{Hybrid approaches}
Combining the stochastic and deterministic approaches has the potential to benefit from efficiency of stochastic and realism of deterministic approaches. 
In an \gls{isac} context, the hybrid models would use a deterministic method to identify primary signal propagation paths and a stochastic method to generate additional objects and clusters. 
For example, 3GPP already includes a hybrid channel model where RT is used to find propagation paths, and stochastic clusters are generated afterward~\cite{3gpp.38.901}. For \gls{isac}, it is important to incorporate the object's RCS into the channel model for more realistic simulation of scattered rays, which has significant implications for applications like object recognition. 
Long-run simulations that maintain consistency in time and space are required for an accurate evaluation of these applications. Therefore, the \gls{gscm} needs to incorporate a spatial consistency model, as  noted in~\cite{thoma2022characterization}, to meet this requirement and avoid inconsistencies and artifacts in the Doppler spectrum.



\subsubsection{Characterizing \gls{isac} channels: recent measurements and modeling at \gls{mmwave} bands}

\begin{figure}[!t]
	\centering
	\includegraphics[width=\columnwidth]{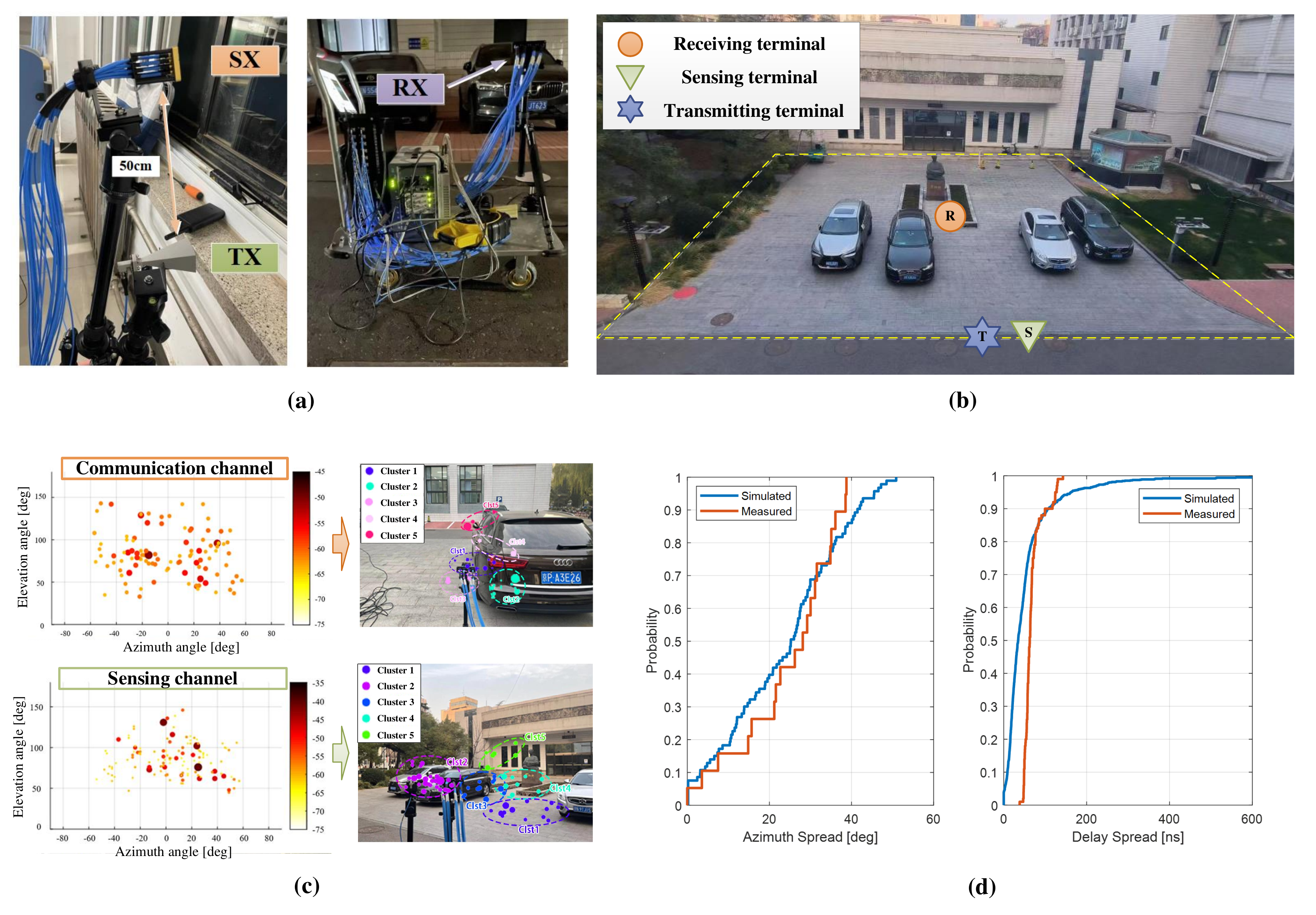}
	\caption{\gls{isac} channel measurement and modeling in \cite{zhang2023mmwave,zhang2022general}. (a) Photos of \gls{isac} channel measurement systems; (b) \gls{isac} channel measurement campaign; (c) Mapping between communication channel and sensing channel; (d) CDF comparisons of azimuth \gls{as} and \gls{ds} of simulated and measured data.}
	\label{ISAC_figure1}
\end{figure}

Recent \gls{isac} channel measurements at 28 GHz frequency band are carried out in Beijing Jiaotong University, China \cite{zhang2023mmwave}. A measurement system is designed as shown in Fig.~\ref{ISAC_figure1}(a), including Tx, sensing terminal (SX) and Rx. Tx and Rx use a directional horn antenna and a 4x8 rectangular antenna array, respectively. SX also uses a 4x8 rectangular antenna array and it is located close to TX to measure sensing channel. The sounding signals have 1 GHz bandwidth and are transmitted with a maximum power of 28 dBm.

Fig.~\ref{ISAC_figure1}(b) shows \gls{isac} channel measurement scenario on campus, and different locations for RX are considered during the measurements. The novel idea of \gls{isac} channel characterization and modeling in \cite{zhang2022general} is to distinguish communication channel and sensing channel from propagation perspective and characterize the both channels jointly in statistical modeling, which accurately reflects the underlying correlation between communication channel and sensing channel. According to mapping relation in environment, some communication clusters are reserved in sensing channels, which are named evolving sensing clusters. Besides, some newly generated clusters only exist in sensing channels, which are named newborn sensing clusters. \gls{isac} channels can be modeled based on the distribution of those clusters. The \gls{isac} channel parameters are estimated using the \gls{sage} algorithm and channel multipaths are clustered using the K-Power-Means algorithm \cite{huang2022artificial1}. Based on measurements, the first five clusters with the maximum powers are mapping and matching to real physical objects in environment as shown in Fig.~\ref{ISAC_figure1}(c). According to the mapping of real physical objects, cluster transition probability from communication channels to sensing channels is firstly defined in \cite{zhang2022general} and statistically modeled. The \gls{ds} and azimuth \gls{as} of simulated and measured \gls{isac} channels are compared in Fig.~\ref{ISAC_figure1}(d) that shows fairly good agreement. This validates that the proposed \gls{isac} channel model has fairly high accuracy.

\subsubsection{Open challenges in \gls{isac} channel modeling}

Channel modeling scenarios for \gls{isac} include 
road and air traffic, logistics, critical infrastructure protection, among others. In such scenarios, \gls{isac} channels cannot be generalised as stationary in a wide sense. Then, models require dynamic scenarios to include moving passive objects, target track motion, pattern interpolation or model-based data compression, and Doppler effects along the track, as well as micro-Doppler, to consider local movements within the target.

\gls{isac} targets cannot be modeled as communication scatterers, and require a model adapted to sensing channels that includes bistatic delay and Doppler, over the current RCS radar approach. Scatterers need 3D geometric modeling, including dynamic state vectors of position and orientation. Correlation between sensing and communication channels in bistatic and distributed scenarios has not been properly studied and modeled so far. On the other hand, mutual interference between the \gls{isac} channels may also be a limiting factor for \gls{isac} applications in some scenarios and system implementations. This knowledge will be relevant for future applications in which the estimation of the communications channel from the sensing one may help saving signalling and radio resources.

For standardisation purposes, it will be worth extending the 3GPP communication models to \gls{isac}, either based on \gls{gscm} or hybrid approaches. Measurements can help verifying if hybrid models preserve the spatial consistency and may be useful for 6G system level evaluation.

\subsection{Channel Measurements and Modeling for ultra large arrays/\gls{mimo}} 

\gls{mimo} technology will continue to evolve for the 6G communication system. It is expected that array with thousands of antenna elements (also refereed to as ultra-large-scale antenna systems,  gigantic \gls{mimo} or extremely large-scale antenna systems) will be accommodated in 6G radios. Radio channel modeling is essential for the system design, optimization and performance evaluation of such ultra-large-scale antenna systems. In this section, state-of-art and key challenges are briefly summarized for radio channel characterization of ultra-large arrays, with a focus on channel sounder design, radio channel parameter extraction and channel modeling. 

Geometry based stochastic channel models, e.g. \gls{3gpp} 25.996 \cite{3GPP_25_996} and \gls{3gpp} 38.901~\cite{3gpp.38.901} are selected as standard channel models for 4G and 5G communication systems, respectively. The \glspl{ue} are small in size and far-from the scatterers and \glspl{bs}. Therefore, plane wave model and stationary channel are typically adopted in the standard channel models. However, these assumptions might be violated for ultra-large-scale \gls{mimo} systems. The large array aperture would require a large far-field distance, which will be violated in practical deployment scenarios. As a result, \glspl{ue} might be located in the near-field region of the BSs. Another effect introduced by the ultra-large-scale \gls{mimo} is the spatial non-stationarity. It has been generally assumed in the standard channel models that the multipath components seen by the array elements are unchanged (i.e. spatially stationary) across array elements. However, spatial non-stationary, i.e. different channels can be observed by different array elements, might exist as the array gets larger. Channel  non-stationary properties have been considered in COST 2100 channel models, where the concept of visibility region was proposed to model spatial non-stationarity over arrays.  As for deterministic channel modeling approach, it is favorable for characterizing site-specific scenarios.  \gls{rt} simulation can in principle well capture the channel characteristics of ultra-massive \gls{mimo} systems, and it is a promising solution for such systems. However, it is computationally heavy to obtain the \gls{rt} channels for all the antenna elements within the large-scale array. The problem will become much more pronounced for  large-scale deployment scenario with many objects, e.g. urban environments. Several alternative approaches have been discussed to reduce the computation complexity, e.g. the database and ray interaction simplification in the METIS model.  

Reliable channel sounders are essential for obtaining high-fidelity channel measurement data. As for ultra-massive \gls{mimo} antenna systems, the focus has been on measuring channel spatial profiles, since the key task of extremely large-scale antenna systems is to better exploit the spatial property of the radio channels. Typical solutions for measuring channel spatial profile reported in the literature include real antenna array (i.e. with parallel RF chains), switched antenna array (i.e. with one RF chain connected to multi-antennas with a switch), phased array, and virtual antenna array. There  exists a trade-off between channel sounder capability and cost. Real antenna array based channel sounder is capable of capturing real-time channel responses, enabling measurements in highly dynamic scenarios. However, its cost and complexity is rather high, especially for ultra-massive antenna systems. Virtual array solution, which has been widely employed already for large-scale antenna based channel sounding, on the other hand, can easily achieve scalable antenna array configuration, making it highly suitable for large-scale antenna based channel measurements. However, it is limited to static scenario and it requires highly accurate positioning and phase coherent measurement system. 

Generic channel parameter estimator, which can accurately extract multipath parameters with high resolution, is highly desirable. Many channel parameter estimators have been reported in the literature. Plane wave assumption is typically adopted to reduce the model complexity. However, this assumption is challenged as the antenna array size gets larger and cell size gets smaller. Narrowband is also assumed in many algorithms, to reduce complexity in multi-domain parameter estimation. However, ultra-wideband system implementation might be expected for the future radio systems, especially at the mmWave and sub-\gls{thz} frequency bands. Another key general assumption in channel parameter estimation is the stationary channel for antenna array elements. This assumption is valid for small-scale arrays as well. However, as the array dimension gets large, this assumption will be eventually violated. As a result, elements across the large-scale array will experience multipath components with different parameters. These observations for ultra-massive \gls{mimo} systems, if not properly considered, will eventually impact the channel parameter estimator performance. Oversimplification in the model will also introduce model mismatch errors for the parameter estimation. 

\subsection{Application of \gls{ml} and \gls{dl} for Propagation Classification, Clustering and Regression} 

An accurate wireless channel model is a necessity to support environment-aware communications. Wireless channel characteristics are vital in stochastic channel modeling (SCM), localization systems, and \gls{ofdm} technology. They are also regarded as key indicators for quality of communication\cite{RN8}. Wireless channel characteristics can be extracted from measurement data or simulation (such as \gls{rt}). However, it is challenging to conduct measurements. Simulation is time-consuming and expensive, especially for complex environments in high-frequency band\cite{10001700}. 

The recent surge of \gls{ai} is revolutionizing almost every branch of science and technology, including wireless channel modeling\cite{9713745}. Many researchers are trying to utilize \gls{dl} models and \gls{ml} algorithms to estimate or generate wireless channel characteristics. 

In the general framework of wireless systems and communications, \gls{ml}/\gls{dl} can be leveraged to address three major problems:
\begin{itemize}
    \item \emph{classification}, e.g. for \gls{los}/N\gls{los} identification \cite{dicicco2023},\cite{MLforV2V}. Reliable and fast detection of \gls{los} can be helpful to assist beamforming techniques or to deploy Fixed Wireless Access networks, as their effectiveness improves in presence of visibility between the wireless devices. Moreover, \gls{los}/N\gls{los} detection can be also beneficial in mobile channel modeling, as different formulas can be effectively applied depending on whether \gls{los} or N\gls{los} conditions occur.
    \item \emph{clustering}, e.g. to identify and group multipath contributions with similar features \cite{he2018clustering,huang2022artificial1}. Multipath clustering is crucial to limit the complexity of channel modeling while catching the essence of the propagation process at the same time.
    \item \emph{regression}, i.e. to get the cause and effect relationship between some propagation markers (like received signal strength, path loss, spread coefficients of the channel, etc.) and some input features relevant to the propagation process. 
\end{itemize}

\begin{figure}
\begin{center}    
  \includegraphics[width=\columnwidth]{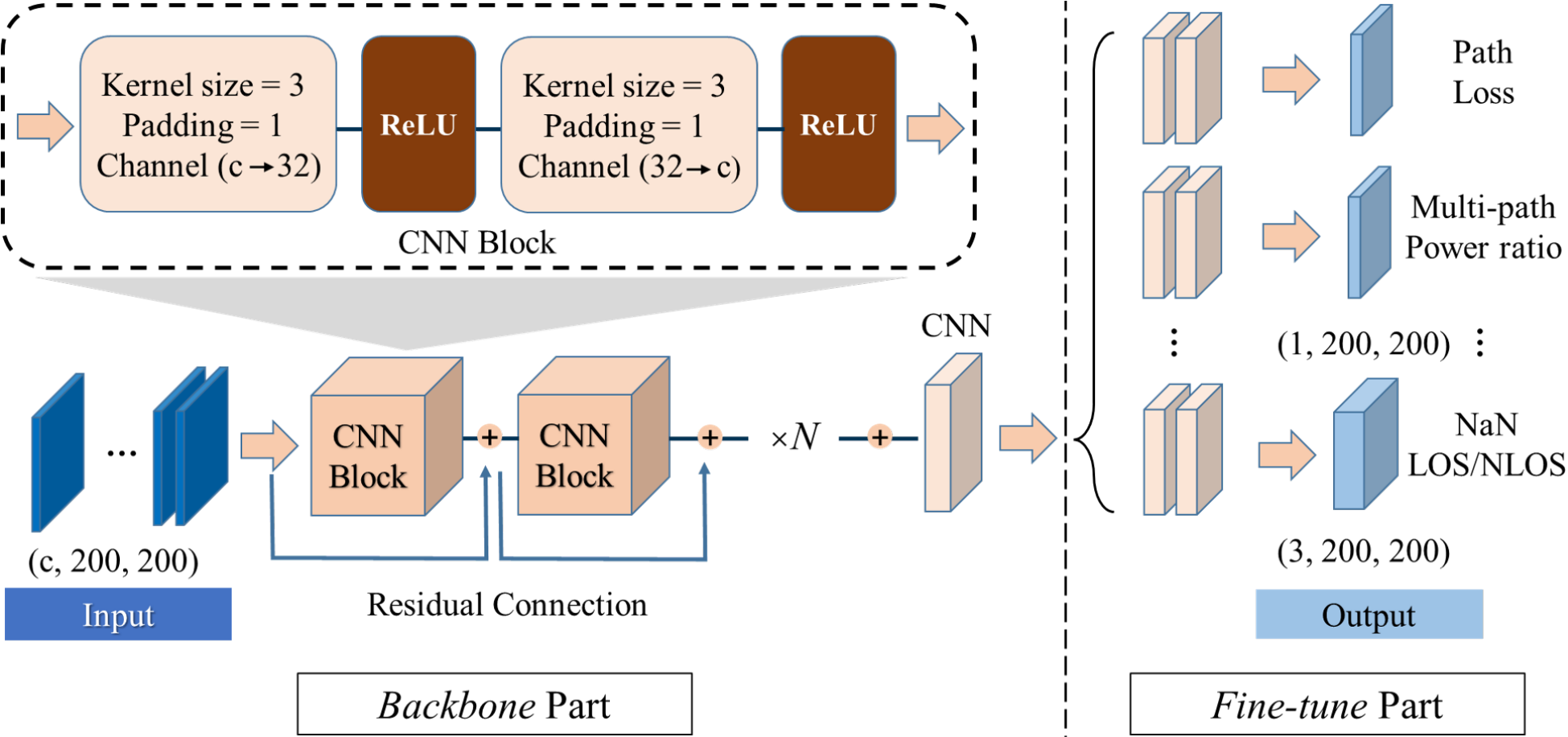}
\end{center}\vspace{-0.5cm}
\caption{The MTL DL model for channel characteristics SR in \cite{10001700}.}
\label{MTLmodel}  
\end{figure}

In \cite{RN58}, authors used linear models, artificial neural networks (ANN), and k-nearest neighbor (KNN) to estimate the power of received radio signals in urban areas. \gls{dl} models are becoming increasingly popular due to their powerful ability in non-linear approximation and massive data processing. \cite{RN26} introduced a deep \gls{cnn} for radio map estimation. Besides channel characteristics in the power domain, many researchers also focused on the temporal domain channel characteristics such as \gls{ds} \cite{IOTJ}\cite{TAP}. Typically, for data (such as images) processed by \gls{dl} models, pixels of different channels are entirely independent. However, channel characteristics (in power, temporal, and angular domain) are correlated because they originate from the same radio propagation process. Therefore, multi-task learning (MTL) can be beneficial in generating multiple channel characteristics simultaneously. \cite{10001700} introduced an MTL \gls{dl} model (Figure \ref{MTLmodel}) for super-resolution (SR) of six kinds of channel characteristics. High-resolution channel characteristics data can be recovered by the proposed MTL \gls{dl} model with low-resolution data as input. The authors also evaluated other mainstream \gls{dl} models. The results indicate that without adjustment, popular \gls{dl} models (such as ResNet, ViT and GAN) can not be applied for SR of wireless channel characteristics. 

Certainly, \gls{ai} will play an increasingly more influential role in channel modeling. Here we propose two suggestions for future research:
\begin{itemize}
\item[(1)] Data is always regarded as the impetus of \gls{ml} models. Measurement data and data by simulation are two primary sources of training data for \gls{ml} to propagation. Novel channel measurement technologies should be studied to reduce the measurement cost and improve the precision of measurement data (such as denoising). Current simulation software should be modified to cater mainstream data formats for popular \gls{dl} computing frameworks such as PyTorch and TensorFlow.

\item[(2)] Compared with typical image and voice datasets, the size of channel characteristics data is relatively small. Lightweight \gls{dl} models are recommended because large \gls{dl} models are inclined to overfit the small training dataset. Another advantage of the lightweight \gls{dl} model is that they are easier to be computed in \glspl{bs} without GPUs. Considering distributed learning and federated learning, the communication cost can be reduced due to fewer parameters being updated.
\end{itemize}

\subsection{Data-Driven Radio Channel Prediction - Extrapolation in Fre\-quen\-cy/Ti\-me-Spa\-ce Domains and for Different Scenes and Systems} 

Channel prediction is to interpolate and extrapolate the channel responses or its properties in frequency and/or spatial-temporal domains. Environment-aware channel prediction also includes the prediction of channels in multiple scenarios/environments. 
Channel predictions have conventionally been dominated by model-driven approaches that reply on physics (e.g., utilizing the closed-form translation and rotation of spherical waves expanded from the received field at antennas \cite{9367515}), mathematics (e.g., using Fourier transform to interpolate channel in frequency domain, using anomaly detection for blockage-aware channel prediction \cite{10001163}) and statistics (e.g., scenario-based stochastic channel models with parameterization). However, with the development of 6G, it is expected to map the physical and virtual worlds and expand the boundaries of human-machine-things connectivity; thus it is expected that channel modeling/predictions could cover all spectra, different systems, and full applications in various scenarios. 
Model-driven prediction approaches could be enhanced with the support of the advanced data-driven methods to expand further the predictable boundaries.

The data-driven methods, or in other words, the \gls{ai} or the \gls{ml}/\gls{dl} methods, are represented by deep neural networks and have been widely used in many fields because of the excellent nonlinear modeling ability. AI-enabled channel modeling/prediction as proposed in \cite{yang2023AI,huang2022artificial2, 10001404} is a disruptive technology, which can well improve the intelligence and accuracy of radio propagation prediction and simulation. Compared with the traditional methods, advantages of AI-enabled channel modeling/prediction are as follows:
\begin{itemize}
\item[(1)] Efficient massive data mining and processing ability: With the explosive growth of available data due to the expansion of applicable bands and scenarios as well as system resolutions and capabilities, the acquisition, storage and processing of massive data have brought significant challenges to traditional channel modeling/prediction methods. Deep neural networks are good at mining complex features in highly dimensional and highly redundant data, and do not rely on additional manual feature screening.
\item[(2)] Strong modeling and adaptive ability: Deep neural network has good performance in nonlinear system modeling. Since neural network can automatically extract input features and establish mapping, it has excellent adaptability and generalization ability when input features change.
\item[(3)] Excellent learning and prediction ability: AI-driven channel model directly learns features of data sets and extracts core factors that have impacts on channels, therefore the predicted outputs can be more essentially derived from the changes of the input features, thus improving accuracy of channel prediction.
\end{itemize}

The \gls{ai} based data-driven methods, good at establishing the relation between massive radio channel data and complex physical propagation environment \cite{he2022guest,mi2023cluster}, and have many potential applicable domains in channel prediction as shown in Fig.~\ref{AI_figure1} and below.
\begin{itemize}
\item[(1)] Scenario-to-Channel: Channel models based on \gls{ai} take scenario features as input and then output channel parameters. The expected model is a group of trained networks by using massive channel and environment data (e.g., \gls{3d} Lidar data and 2D image data \cite{mi2023cluster, 10001404}). The trained deep neural network is expected to well establish the mapping relationship from scenario to channel (e.g., path loss, major cluster), and the goal is to predict radio frequency channel parameters from other modalities of image or/and point clouds.

\begin{figure}[!t]
	\centering
	\includegraphics[width=\textwidth]{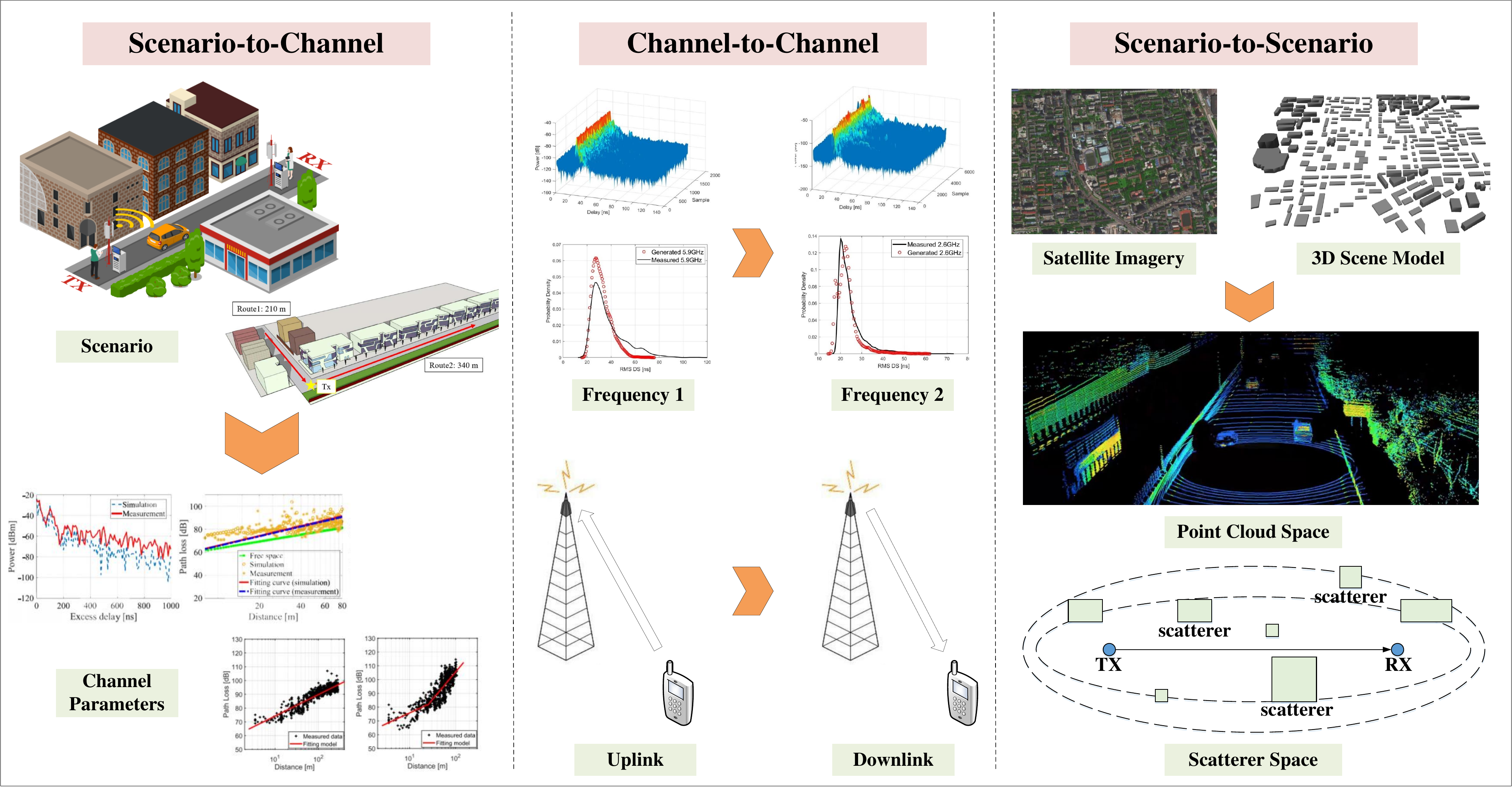}
	\caption{Implementation examples of \gls{ai} for channel modeling.}
	\label{AI_figure1}
\end{figure}

\item[(2)] Channel-to-Channel: Typical application of \gls{ai} enabled channel-to-channel mapping is data enhancement and frequency migration including both interpolation and extrapolation. Neural networks can be used to learn limited channel data and quickly generate massive data with similar propagation characteristic; it is valuable for the applications requiring massive channel data, e.g., in over-the-air wireless device testing chambers \cite{8657705}. Furthermore, \gls{ai} models can learn frequency impacts on different channels and predict the channel for unknown frequencies. A typical case is to use uplink channels to predict downlink channels in Frequency Division Duplex systems.

\item[(3)] Scenario-to-Scenario: Mapping from physical environment to electromagnetic virtual environment is a potential field, and it's essentially to reconstruct environment from the perspective of radio wave propagation. Scenario-to-scenario mapping aims to construct a scenario model of point cloud or scatterer space using massive data such as satellite images. The reconstructed scenario model contains rich electromagnetic environment information, and it can not only serve to predict channels but also to realize electromagnetic environment perception.
\end{itemize}

In addition to the above-mentioned deep learning based data-driven channel modeling/predictions, there are also other complexity/computation less-intensive methods that can be explored to be used for channel prediction. For instance, the evolutionary algorithms (EA) or generic algorithms (GA) \cite{9767600, 9410934, 9743500}. Inspired by biological evolution, the EA algorithms encode candidate solutions using chromosomes and provides a fitness function determining their qualities; over iterations, crossover and mutation are performed to generate new chromosomes and selection is effectuated to preserve good chromosomes. EA or GA could perform channel prediction with the assistance of \gls{rt} tools for fitness function and reach near-optimal solutions/predictions. 

Most above mentioned methods predict channel in spatial-temporal (environment and mobility) and frequency domains. To expand the boundaries for future applications, the channel prediction needs to be performed across different systems as well. As such, the data-driven algorithms need to be trained by using data captured in different domains and systems (e.g., radar monostatic backscatterd channel data and communication signal bistatic channel data, \cite{10000833, lirias4005390}). Training with data captured by different systems could make the trained AI-based algorithm robust to different system setups and non-linearities, and the goal is to obtain the domain and system invariant channel prediction algorithm.

\subsection{Hardware-in-the-Loop Radio Channel Emulation} 
The \gls{3gpp} generally defines many standard channel models for wireless communication system simulation and evaluation. However, theoretical channel model cannot be directly used for hardware system and terminal simulation/evaluation such as air interface testing. Radio channel emulator can act as a representation of the real-world radio channel, and it enables creation of mathematical channel models representing physical radio signal transmission \cite{eslami2009design,fan2018flexible}. It is still challenging for channel emulation in complex environments such as high-mobility scenario, massive \gls{mimo} scenario, etc.

Wireless channel emulator uses a down-converter to transform RF signal to baseband, and then uses high-speed digital signal processors such as FPGA to achieve digital filtering of baseband signals based on channel models, generating a signal that incorporates channel effects. Finally, RF output of the signal is achieved through an up-converter. Therefore, channel emulator can achieve equivalent substitution for field measurements in laboratory, spanning various stages of wireless communication research, core equipment development, network planning optimization, and network operation. Many researches have been conducted by COST INTERACT to improve channel emulation.

\begin{figure}[!t]
	\centering
	\includegraphics[width=\columnwidth]{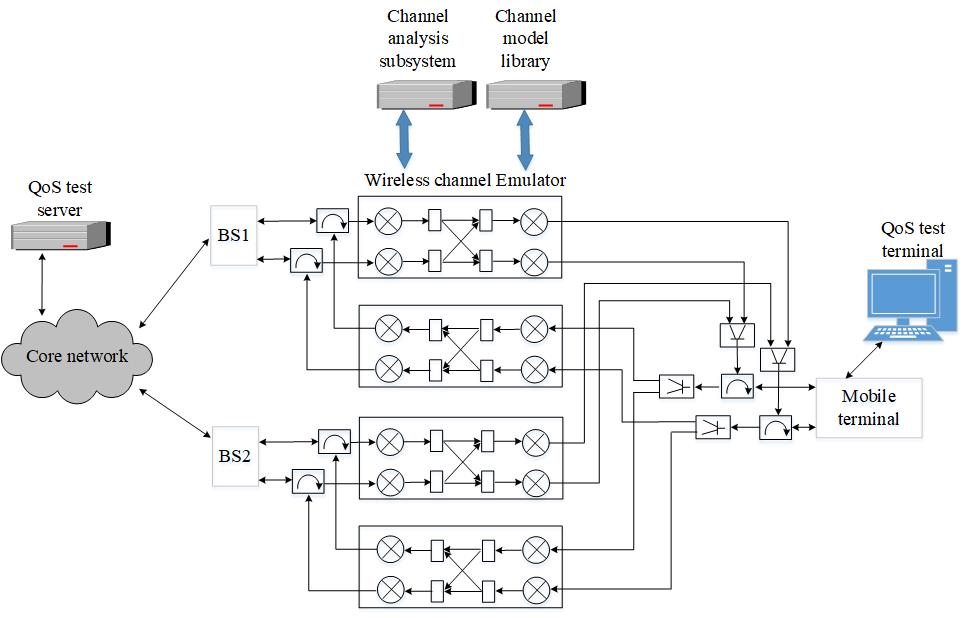}
	\caption{Hardware-in-the-loop emulation system structure for high-mobility communication.}
	\label{Emulation_figure0}
\end{figure}

\begin{figure}[!t]
	\centering
	\includegraphics[width=\columnwidth]{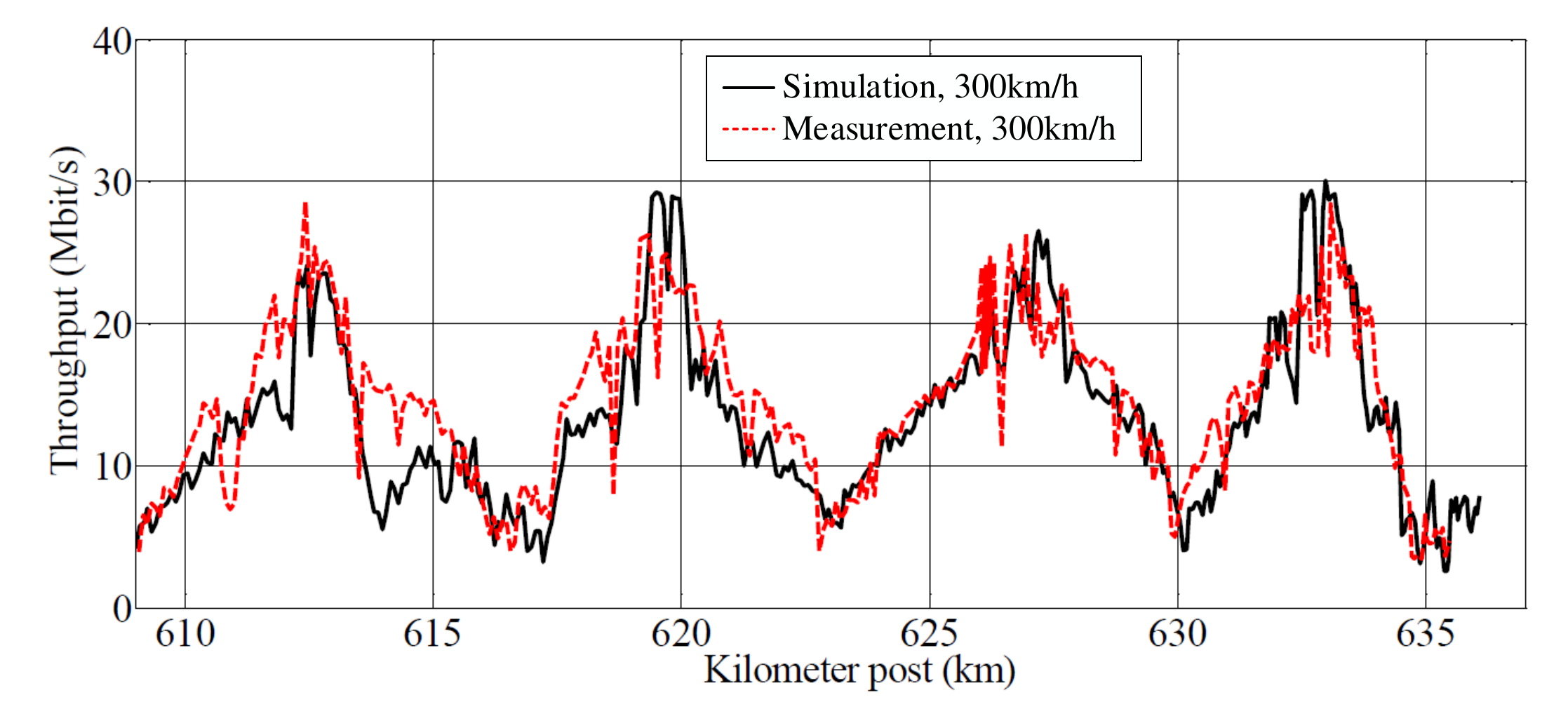}
	\caption{Example plots of instantaneous downlink throughput.}
	\label{Emulation_figure1}
\end{figure}

\begin{figure}[!t]
	\centering
	\includegraphics[height=6.9cm,width=7.5cm]{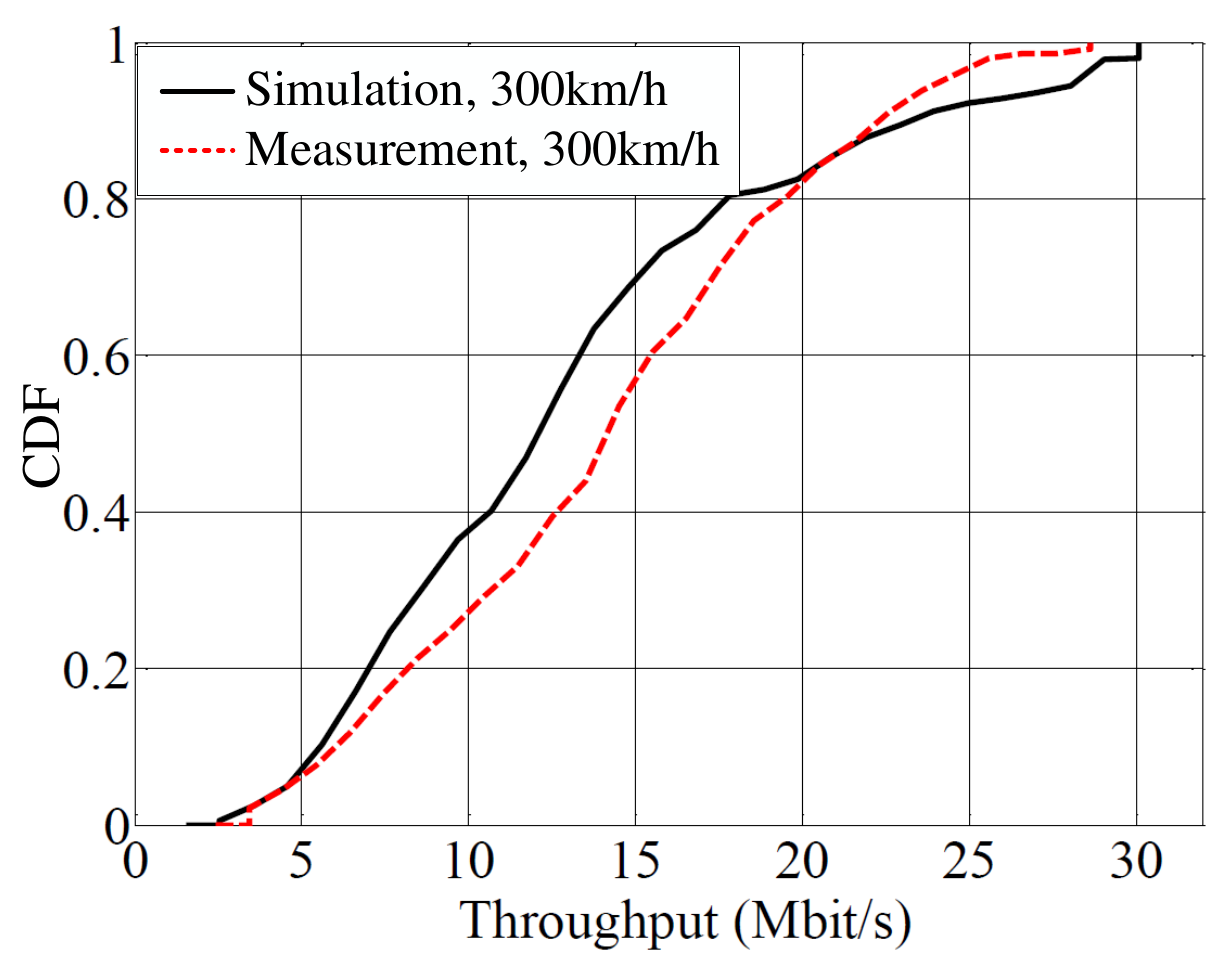}
	\caption{Example plots of downlink throughput distribution.}
	\label{Emulation_figure2}
\end{figure}

In order to further improve channel emulation especially for high-mobility scenario, a novel  hardware-in-the-loop channel emulator is firstly developed by Beijing Jiaotong University so that real-world high-mobility radio environment can be accurately modeled and physically implemented \cite{fei2017development,ding2015research,fei2013hardware}, and the architecture is shown in Fig.~\ref{Emulation_figure0}. The emulation system uses two \glspl{bs} and one core network, and each BS is configured as downlink 2×2 \gls{mimo} space division multiplexing mode. In order to evaluate network performance, a QoS test server and test terminal are set up on core network side and mobile terminal side, respectively. In order to verify accuracy of the emulation system in high-mobility scenario, we use test instrument and network consistent with high-mobility field test to evaluate RSRP, SINR, downlink throughput by the hardware-in-the-loop channel emulator, and compare with field test results. Here, we consider high-speed railway scenario with 300 km/h moving speed. A large body of wideband channel measurements at 450 MHz, 900 MHz, 2.1 GHz bands are conducted along “Beijing-Shanghai”, “Beijing-Tianjin”, and “Beijing-Shenyang” high-speed railway lines, and measurement-based channel models are developed for high-mobility channel emulation \cite{ding2018,zhang2018measurement}. The instantaneous downlink throughput values and CDF curve through hardware-in-the-loop emulation and onboard measurements are compared in Fig.~\ref{Emulation_figure1} and Fig.~\ref{Emulation_figure2}. It is found that the simulated downlink throughput by using the hardware-in-the-loop channel emulator is consistent with measurements and prediction error is less than 10\% \cite{ding2018}.

Currently, channel emulation, especially in high-mobility scenarios, mainly faces the following challenges: i) Conducting measurements in high-mobility scenario is difficult, lacking joint validation with application-level transmission performance; ii) Most channel emulators adopt an instrument-based architecture, which has limited computational and storage capabilities. This limitation hinders the generation of large-scale channel coefficient matrices in high-mobility scenario; iii) Effective emulation time is short, dynamic emulation capabilities are limited, frequency range and bandwidth are restricted, and there is also a lack of emulation capabilities for super-large-scale antenna arrays.

\subsection{Channel sensing using advanced antenna concepts for \gls{mmwave} and beyond } 

\gls{mmwave} and (sub-)\gls{thz} bands open unprecedented opportunities for environment-aware communications. The available spectrum in these frequency ranges enables ultra-high data rates and high-accuracy sensing applications, which are needed for the use cases considered in 6G~\cite{one6G, tong20216G}. At such frequencies, highly directional beams are used to mitigate large free-space attenuation, which in turns enables better exploitation of spatial resources. Sensing the channel in its angular dimension is therefore of utmost importance but becomes challenging as carrier frequency gets higher.

Channel characterization that performs directional measurements in mmWave and (sub-)\gls{thz} bands typically uses rotating horn antennas to benefit from antenna gain and thus increases measurement dynamic (see~\cite{Priebe13, Takahashi22} for indoor scenario and~\cite{Abbasi20} for outdoor scenario). However, steering narrower antenna beams across azimuth and elevation at both transmitter and receiver leads to prohibitive measurement duration. To decrease it, the study in~\cite{Guzman22}, conducted within the COST INTERACT action, implements a measurement based ray-launcher to estimate the double-directional path data from single-directional radio channel sounding. Other channel sounders use antenna arrays to avoid any mechanical displacement and characterize the channel faster. This includes classical phased arrays~\cite{Bas19} or switched arrays~\cite{Wang19} and lens-based arrays~\cite{Sayeed16} to decrease cost and hardware complexity.

While these approaches are suitable for channel characterization, their high cost and complexity limit their applicability for sensing in actual communications such as for beam alignment and handover in mobile scenario or estimating \gls{csi} for \gls{ris}. These use cases require fast and energy efficient techniques to discover the angular properties of the channel. 

An alternative approach is to use a dedicated peculiar antenna that estimates directions of arrival (DoA) with a single radio frequency chain. This results in a non-expensive system with real-time sensing capabilities. Such solutions leverage the frequency diversity that inherently exists in the radiation pattern of some classes of radiating structures. Those devices are purely passive and exhibit therefore low complexity and easy calibration procedure. Cavity-backed metasurfaces~\cite{Abbasi20b}, lens-loaded cavities~\cite{Yurduseven19}, or leaky-wave antennas (LWA)~\cite{Sarrazin22, Sarrazin23} exhibit such properties, with the latter being a lower profile solution. LWAs also exhibit a tractable beam scanning behavior with frequency~\cite{Oliner19} which enables using standard DoA estimation techniques such as monopulse-based~\cite{Poveda2019} or MUSIC algorithm~\cite{Sarrazin21}. While LWAs represent a cost-effective solution to estimate DoA at mmWave and (sub-)\gls{thz} frequencies, they typically need to operate over a large frequency bandwidth in order to scan a large field of view (FoV), which makes them unpractical for most communication standards. 

This issue has been tackled in the literature with different approaches. At \gls{mmwave}, the LWA scanning velocity has been improved by loading the leaky guiding structure with a dense metasurface~\cite{Zhang19} or by adding an extra dispersive lens \cite{Oscar22}. However, the required bandwidth to scan a large FoV remains larger than typical frequency channels used in telecommunications. The works in~\cite{Emara20, Poveda20} exploit several multiport LWAs while \cite{Paaso13} uses reconfigurable LWAs. These approaches achieve\st{s} AoA estimation over a large FoV at the expense of cost and/or complexity.

\begin{figure}[htpb]
	\centering
	\includegraphics[width=.85\columnwidth]{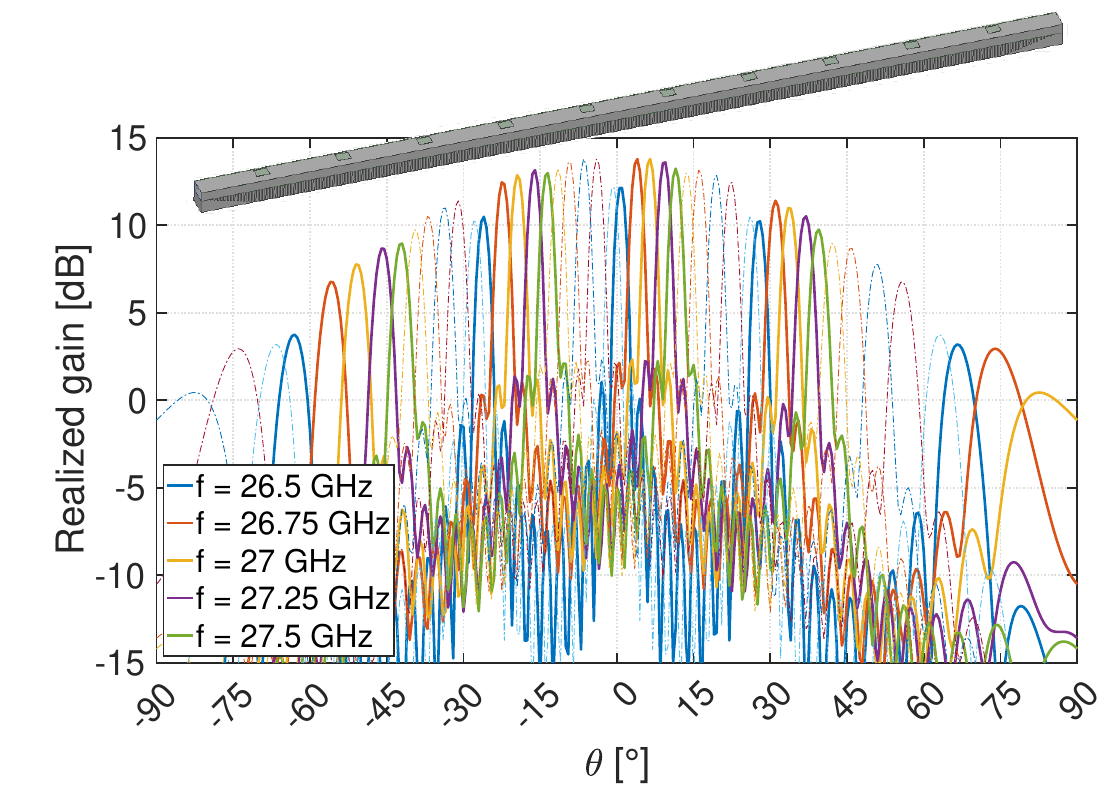}
	\caption{Leaky-wave antenna radiation pattern showing the multiple-beam scanning in the 26.5-27.5 GHz band (solid lines represents the pattern generated by one LWA port and dotted lines the patter generated by the other port).}
	\label{LWApattern}
\end{figure}

\begin{figure}[htpb]
	\centering
	\includegraphics[width=.85\columnwidth]{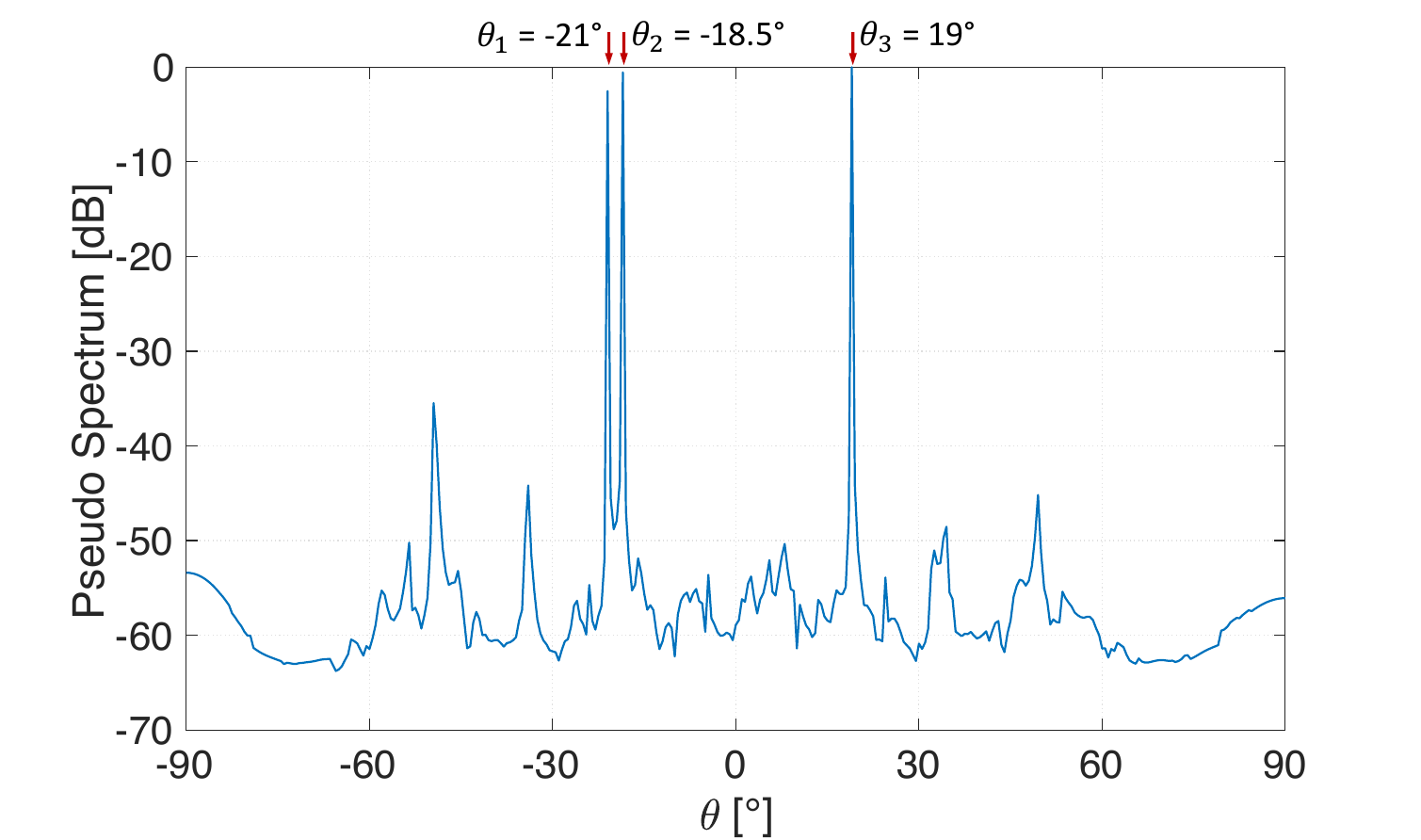}
	\caption{MUSIC pseudo spectrum obtained with the dual-port LWA (100 snapshots, SNR = 10 dB, 144 subcarriers, 3 uncorrelated sources of DoA = $\theta_2$, $\theta_2$, $\theta_3$).}
	\label{LWApseudospectrum}
\end{figure}

Recently, it was proposed in~\cite{Sarrazin21} to exploit LWAs able to radiate multiple beams at each frequency. This multibeam operation is achieved by increasing the period of the spatial modulation of periodic LWAs. This generates multiple fast spatial harmonics, each one contributing to a beam in the far-field radiation. In doing so, the FoV, at each frequency, is divided by the number of beams, which in turns greatly reduces the bandwidth  required for a single beam to scan its angular sub-region. \cite{Sarrazin21} shows that a subspace-based algorithm such as MUSIC can distinguish the DoAs of incoming sources among the multiple beams. Two proofs of concept have been developed within COST INTERACT in the 28~GHz band: an E-plane scanning single-port LWA~\cite{Sarrazin22} which needs a 2-GHz-bandwidth to scan the whole FoV (i.e., 7.4\% fractional bandwidth) and an H-plane scanning dual-port LWA~\cite{Sarrazin23} requiring only 1-GHz-bandwidth to scan the whole FoV (i.e., 3.7\% fractional bandwidth). The former was designed in a substrate-integrated waveguide technology while the latter is a fully metallic structure, which does not suffer from dielectric losses and therefore increases radiation efficiency. The H-plane scanning dual-port LWA is based on a corrugated waveguide modulated by rectangular slots. Its geometry is shown in Fig.~\ref{LWApattern} along with its radiation pattern. Up two five beams at each frequency can be observed whose directions steer with frequency. The MUSIC pseudo spectrum in Fig.~\ref{LWApseudospectrum} shows that the three DoAs considered in this example are well retrieved with no ambiguity among the multiple beams.

 Currently, channel sensing using such advanced antennas concepts still faces some challenges. First, angular estimation based on frequency diversity is prone to frequency fading for spatially non-resolvable multipath components. Consequently, techniques are to be investigated to improve channel sensing robustness. Second, sensing in both elevation and azimuth planes is yet to be done. Therefore, 2D scanning with a single LWA is also an exciting perspective of research, which could lead to cost-effective solution for sensing in future mobile generations. Finally, future works should be carried out to make use of such advanced antennas in beam management to ultimately extend the time-frequency 2D resource grid used so far in 5G to a space-time-frequency 3D grid, enabling a seamless exploitation of the beam space.


\newpage
\section{Conclusions and future outlook}
\label{sec:Conclusions}
\begin{itemize}
    \item[]\textbf{by Vittorio Degli Esposti}
\end{itemize}

The purpose of this paper is to summarize the key challenges in the field of radio channel measurement and modeling that need to be addressed to support the development of next generation (6G) wireless networks. Additionally, it aims at providing an overview of the main research activities undertaken by the scientific community, particularly within  COST CA20120 "INTERACT", in order to achieve those goals.
Next generation wireless networks will have to deal with a variety of environments and applications, with frequencies ranging from sub-6~GHz to THz, from sparse to ultra-dense networks, ultra-high performance links, and including sensing, imaging, and smart-environment applications. Therefore, a number of new studies are required to address relevant issues such as material and propagation characteristics at the new frequency bands, more sophisticated channel sounding techniques and novel modeling methodologies, including the use of machine learning techniques.

Section \ref{sec:Framework} introduces the study of propagation mechanisms and parameters that's fundamental for the definition of channel measurement and modeling techniques. Making use of proper measurement setups, several experimental studies are addressing the analysis of basic propagation mechanisms, such as the increased wall-penetration loss with frequency and blockage loss from humans and objects. Furthermore, several studies address measurement and modeling of diffuse scattering from surfaces and due to material variability, with a focus on polarization characteristics and the enforcement of reciprocity in directional scattering models. The recent advent of \gls{ris} has spurred research on the effect of such surfaces on propagation, with particular emphasis on the comparison between electromagnetic simulation and simplified scattering models, the power-decay trend of reflection from such surfaces and the development of macroscopic modeling approaches.

Other studies are addressing higher-level propagation characteristics, such as fading correlation over space and frequency. Some reports show evidence that power spectrum shapes do not change noticeably across different frequencies, while there is indication from comparative channel sounding that the channel becomes more “sparse” at higher frequencies, at least up to sub-THz frequencies, with a lower degree of multipath richness.

An important research activity within the COST INTERACT community is focused on channel sounding at both sub-6 GHz frequencies and above. Channel sounding techniques are described in Section~\ref{sec:Sounder} while channel measurement results are summarized in Section~\ref{Sec:sounding}.

Measurement campaigns in sub-6 GHz bands have been conducted to characterize wireless propagation, with particular focus on vehicular scenarios. Researchers are now exploring slightly higher frequency ranges, such as those in the so-called mid-band, or FR3 range (7-24 GHz) that is under the spotlight to overcome spectrum congestion and support massive \gls{mimo} systems. Different techniques and array configurations were employed to estimate multipath components and study non-stationarity among antenna elements.

Several other studies have addressed propagation and channel sounding techniques at \gls{mmwave} and sub-THz frequencies. To compensate for the higher path-loss, high-gain antenna systems are needed, making spatial channel characteristics crucial for system design and this fact is reflected in channel sounding techniques. Since reliable massive-\gls{mimo} channel sounders are still unavailable at these frequencies, several studies resort to virtual array techniques, which require a great deal of measurement time. One crucial issue is therefore the reduction of measurement time. Another important activity within COST INTERACT is the collection of channel measurements into a unitary database that should also include detailed information on the measurement environment and technique and can be used for channel modeling and simulation purposes, as described in Section \ref{sec:Measurements}.

On the channel modeling side, a great deal of activity is being carried out, with a focus on Geometrical Stochastic Channel Models (GSCM), map-based models, Machine Learning (ML) based approaches, and advanced ray tracing techniques, as described in Section \ref{sec:Methodologies}. GSCM models are widely used for simulation of wireless systems, considering scatterers in the environment and modeling signal propagation through multiple paths. Various works have proposed GSCM models for different use cases such as vehicular and rail communications. Map-based models use simplified maps of the environment to capture spatial consistency among different links. ML-based approaches have gained attention for wireless channel characterization, where ML algorithms can improve the accuracy of propagation models or provide a black-box representation of the channel, albeit with the drawback of a time-consuming and critical training phase. Tabular data is well-suited for ensemble models like Random Forests and gradient boosting decision tree models, while space-related, unstructured data (e.g., images) can be effectively processed using deep learning models such as convolutional neural networks and appear quite attractive for propagation modeling, given its intrinsic spatial characteristics. Ray tracing can be used as a low-cost alternative to measurements for the training phase. At the same time, advanced ray tracing techniques including parallelization techniques, dynamic ray tracing techniques and ray-based techniques for reducing the computational burden for ultra-large arrays and reconfigurable intelligent surfaces are also being developed. A range of relevant techniques  that are used for parameter estimation and clustering algorithms are covered in Section~\ref{sec:Estimation}, where it is shown there is a strong research ongoing on both the  ``classical'' and learning-based techniques.

Finally, a look into new technologies in the field of channel measurement and modeling is given in Section \ref{sec:NewTechnologies}. 
Among the novel channel modeling techniques that are being addressed, we can mention anticipative channel prediction for dynamic scenarios based on Artificial Intelligence (AI) techniques. AI-based methods have applications in scenario-to-channel prediction, channel-to-channel mapping for data enhancement and frequency migration, and scenario-to-scenario mapping to reconstruct the electromagnetic environment. Hardware-in-the-loop channel emulators are also being developed to accurately model and physically implement real-world high-mobility radio environments. Finally, novel leaky-wave antennas are being proposed for channel sensing and directional channel measurements without the use of rotating directive antennas or arrays.

All considered, thanks to the foreseen new frequency bands, application scenarios and technology developments, we can conclude that research on radio channel characterization and modeling is as active as ever.

\newpage
\section*{Author affiliations}

\begin{tabularx}{\textwidth}{X X X}
    \textbf{Name} & \textbf{Affiliation} & \textbf{Email}\\\hline

    Andrej Hrovat & Institut ``Jozef Stefan'' & andrej.hrovat@ijs.si\\\hline

    Bo Ai & Beijing Jiaotong University & boai@bjtu.edu.cn \\\hline

    Conor Brennan & Dublin City University &  conor.brennan@dcu.ie\\\hline

    Dan Fei & Beijing Jiaotong University & dfei@bjtu.edu.cn\\\hline
    
    Danping He & Beijing Jiaotong University &  hedanping@bjtu.edu.cn\\\hline
    
    Diego Andres Dupleich & TU Ilmenau & Diego.Dupleich@tu-ilmenau.de\\\hline

    Enrico Maria Vitucci & University of Bologna & enricomaria.vitucci@unibo.it\\\hline

    Franco Fuschini & University of Bologna & franco.fuschini@unibo.it\\\hline

    Guido Valerio &
    Sorbonne University &
    guido.valerio@sorbonne-universite.fr\\\hline

    Joonas Kokkoniemi & Oulu University & joonas.kokkoniemi@oulu.fi\\\hline
    
    Julien Sarrazin & Sorbonne University & julien.sarrazin@sorbonne-universite.fr\\\hline

    Katsuyuki Haneda & Aalto University & katsuyuki.haneda@aalto.fi\\\hline

    Ke Guan & Beijing Jiaotong University &  ke.guan.bjtu@qq.com\\\hline

    Marco Di Renzo & Centrale Supelec &  marco.direnzo@l2s.centralesupelec.fr\\\hline

    Marco Skocaj & University of Bologna & marco.skocaj@unibo.it\\\hline

    Mate Boban & Huawei, Munich Research Center & mate.boban@huawei.com\\\hline

    Mi Yang & Beijing Jiaotong University & myang@bjtu.edu.cn\\\hline

    Narcís Cardona & Valencia Polytechnic University & ncardona@iteam.upv.es \\\hline

    Nicola Di Cicco & Politecnico di Milano & nicola.dicicco@polimi.it\\\hline

    Ruisi He & Beijing Jiaotong University & he.ruisi.china@gmail.com \\\hline

    Tomaz Javornik & Institut ``Jozef Stefan'' & tomaz.javornik@ijs.si\\\hline

    Tommaso Zugno & Huawei, Munich Research Center & tommaso.zugno@huawei.com\\\hline
    
    Vittorio Degli Esposti & University of Bologna & v.degliesposti@unibo.it\\ \hline
    
    Wei Fan & Aalborg University & wfa@es.aau.dk\\ \hline

    Wenfei Yang & Huawei Technologies Co., Ltd &
    yangwenfei4@huawei.com\\\hline

    Xiping Wang & Beijing Jiaotong University & wangxiping@bjtu.edu.cn\\\hline
    
    Xuesong Cai & Lund University &  xuesong.cai@eit.lth.se \\\hline

    Yang Miao & University of Twente & y.miao@utwente.nl\\\hline

\end{tabularx}

\bibliographystyle{IEEEtran}
\bibliography{references,xuesong,references_DD,ref_Katsu,Wei_library}   
\end{document}